\documentclass[pre,twocolumn,showpacs,showkeys,superscriptaddress,preprintnumbers,floatfix]{revtex4-1}

\usepackage{etex}
\usepackage{ifpdf}
\usepackage{hyperref}
\usepackage{dcolumn}
\usepackage{url}
\usepackage{amsmath}
\usepackage[makeroom]{cancel}
\usepackage{amscd}
\usepackage{amsfonts}
\usepackage{amssymb}
\usepackage{bm}   
\usepackage{bbm}
\usepackage{verbatim}
\usepackage{stmaryrd}
\usepackage{amsthm}
\usepackage{xcolor}
\usepackage{setspace}
\usepackage{braket}

\usepackage{diagbox}

\usepackage{tikz}
\usetikzlibrary{arrows}
\usetikzlibrary{automata}
\usetikzlibrary{decorations.pathreplacing}
\usetikzlibrary{positioning}
\usetikzlibrary{plotmarks}
\usetikzlibrary{calc}
\usetikzlibrary{patterns}
\usetikzlibrary{external}
\usepackage{pgfplots}
\pgfplotsset{compat=newest} 

\usepackage{graphicx}

\theoremstyle{plain}    
\theoremstyle{plain}    
\theoremstyle{plain}    
\theoremstyle{plain}    
\theoremstyle{plain}    
\theoremstyle{plain}    
\theoremstyle{plain}    
\theoremstyle{plain}    
\theoremstyle{plain}    
\theoremstyle{plain}    
\theoremstyle{plain}    
\theoremstyle{plain}

\newcommand{\eM}     {\mbox{$\epsilon$-machine}}
\newcommand{\eMs}    {\mbox{$\epsilon$-machines}}

\newcommand{\eT}     {\mbox{$\epsilon$-transducer}}
\newcommand{\eTs}    {\mbox{$\epsilon$-transducers}}

\newcommand{\MeasAlphabet}  {\mathcal{A}}

\newcommand{\CausalState}   { \mathcal{S} }

\newcommand{\Cmu}       {C_\mu}
\newcommand{\hmu}       {h_\mu}

\newcommand{\ProcessAlphabet}   {\MeasAlphabet}

\newcommand{\forward}{+}
\newcommand{\reverse}{-}
\newcommand{\forwardreverse}{\pm} 

\newcommand{\FutureCausalState} { {\CausalState}^{\forward} }

\newcommand{\PastCausalState}   { {\CausalState}^{\reverse} }

\newcommand{\lastindex}[2]{
  \edef\tempa{0}
  \edef\tempb{#2}
  \ifx\tempa\tempb
    \edef\tempc{#1}
  \else
    \edef\tempa{0}
    \edef\tempb{#1}
    \ifx\tempa\tempb
      \edef\tempc{#2}
    \else
      \edef\tempc{#1+#2}
    \fi
  \fi
  \tempc
}

\newcommand{\I}{\mathbf{I}}

\newcommand{\CSjoint}[1][,]{
   \edef\tempa{:}
   \edef\tempb{#1}
   \ifx\tempa\tempb
      \ensuremath{\FutureCausalState\!#1\PastCausalState}
   \else
      \ensuremath{\FutureCausalState#1\PastCausalState}
   \fi
}

\newif\ifpm
\edef\tempa{\forwardreverse}
\edef\tempb{\pm}
\ifx\tempa\tempb
   \pmfalse
\else
   \pmtrue
\fi

\renewcommand{\H}{\operatorname{H}}
\renewcommand{\I}{\operatorname{I}}

\newcommand{\kB}{k_\text{B}}  

\begin{document}

\title{Leveraging Environmental Correlations:\\
The Thermodynamics of Requisite Variety}

\author{Alexander B. Boyd}
\email{abboyd@ucdavis.edu}
\affiliation{Complexity Sciences Center and Physics Department,
University of California at Davis, One Shields Avenue, Davis, CA 95616}

\author{Dibyendu Mandal}
\email{dibyendu.mandal@berkeley.edu}
\affiliation{Department of Physics, University of California, Berkeley, CA
94720, U.S.A.}

\author{James P. Crutchfield}
\email{chaos@ucdavis.edu}
\affiliation{Complexity Sciences Center and Physics Department,
University of California at Davis, One Shields Avenue, Davis, CA 95616}

\date{\today}
\bibliographystyle{unsrt}

\begin{abstract}
Key to biological success, the \emph{requisite variety} that confronts an adaptive organism is the set of detectable, accessible, and controllable states in its environment. We analyze its role in the thermodynamic functioning of information ratchets---a form of autonomous Maxwellian Demon capable of exploiting fluctuations in an external information reservoir to harvest useful work from a thermal bath. This establishes a quantitative paradigm for understanding how adaptive agents leverage structured thermal environments for their own thermodynamic benefit. General ratchets behave as memoryful communication channels, interacting with their environment sequentially and storing results to an output. The bulk of thermal ratchets analyzed to date, however, assume memoryless environments that generate input signals without temporal correlations. Employing computational mechanics and a new information-processing Second Law of Thermodynamics (IPSL) we remove these restrictions, analyzing general finite-state ratchets interacting with structured environments that generate correlated input signals. On the one hand, we demonstrate that a ratchet need not have memory to exploit an uncorrelated environment. On the other, and more appropriate to biological adaptation, we show that a ratchet must have memory to most effectively leverage structure and correlation in its environment. The lesson is that to optimally harvest work a ratchet's memory must reflect the input generator's memory. Finally, we investigate achieving the IPSL bounds on the amount of work a ratchet can extract from its environment, discovering that finite-state, optimal ratchets are unable to reach these bounds. In contrast, we show that infinite-state ratchets can go well beyond these bounds by utilizing their own infinite ``negentropy''. We conclude with an outline of the collective thermodynamics of information-ratchet swarms.
\end{abstract}


\preprint{Santa Fe Institute Working Paper 16-09-XXX}
\preprint{arxiv.org:1609.XXXXX [cond-mat.stat-mech]}

\maketitle


\setstretch{1.1}
\section{Introduction}
\label{sec:Introduction}
The mid-twentieth century witnessed an efflorescence in information and control
and, in particular, the roles they play in biological adaptation
\cite{Shan56a}. Norbert Wiener's linear prediction theory \cite{Wien49,Wien81}
and Claude Shannon's mathematical theory of communication
\cite{Shan48a,Shan49a,Shan59a,Shan61a} stood out as the technical
underpinnings. It was Wiener, though, who advocated most directly for a broad
development of a new calculus of control and adaptation, coining the term
``cybernetics'' \cite{Wien48,Wien88a}. The overall vision and new methods of
information theory and linear stochastic processes stimulated a tremendous
enthusiasm and creativity during this period.

It must be said that, despite substantial efforts throughout the 1950s and
1960s to develop ``general systems'' theories and the like
\cite{Bert69a,Ashb60a}, at best, only modest successes transpired which
addressed Wiener's challenges for cybernetics \cite{Quas58a}. Historians of
science claimed, in fact, that progress was inhibited by the political tensions
between the West and East during the Cold War \cite{Conw06a}. More practically,
one cause was the immodest complicatedness of the systems targeted---weather
control, the brain, and social design. In short, there simply were not the
powerful computational and mathematical tools required to understand such
large-scale, complex systems. This all said, we must not forget that the
intellectual fallouts from this period---the development of communication,
coding, computation, and control theories---substantially changed the landscape
of the engineering disciplines and irrevocably modified modern society.

Now, at the beginning of the 21st century, it seems time to revisit the broad and ambitious goals these early pioneers laid out. For, indeed, the challenges they introduced are still with us and are evidently starting to reveal dire consequences of our failure to understand the dynamics and emergent properties of large-scale complex systems, both natural and man-made. Optimistically, very recent developments in nonlinear dynamics~\cite{Crut12a} and nonequilibrium thermodynamics~\cite{Klag13a} give hope to finally achieving several of their goals, including reframing them in ways that will facilitate physical implementation. Here, we elucidate cybernetics' Law of Requisite Variety in light of these recent advances.

W. Ross Ashby was one of cybernetics's best expositors \cite{Ashb57a}, having an impact that rivaled Wiener's advocacy. Principle to Ashby's approach was his concept of \emph{requisite variety}. The requisite variety that confronts an adaptive system is the set of accessible, detectable, and controllable states in its environment. In its most elementary form, Ashby re-interpreted Shannon's notion of information-as-surprise, retooling it for broader application to biological and cognitive systems \cite{Ashb60a}. In this, though, he was anticipated by $30$ years by Leo Szilard's successful purging of Maxwell Demon \cite{Szil29a, Leff02a}: ``... a simple inanimate device can achieve the same essential result as would be achieved by the intervention of intelligent beings. We have examined the `biological phenomena' of a nonliving device and have seen that it generates exactly that quantity of entropy which is required by thermodynamics''. In laying out the thermodynamic costs of measurement, and so showing any demon is consistent with the Second Law of Thermodynamics, Szilard not only anticipates by two decades Shannon's quantitative measure of information but also Wiener's conception of cybernetics in which stored information plays a \emph{functional role}.

The conceptual innovation in Szilard's analysis, still largely under
appreciated, is his identifying two distinct kinds of information. On the one
hand, there is surprisal; Shannon's notion that later on lead to an algorithmic
foundation for randomness and probability \cite{Kolm65,Kolm83,Chai66,Vita90a}.
Its parallel in physics is a system's thermodynamic entropy \cite{Jayn57a}.
The Demon monitors statistical fluctuations in its heat-bath environment. On
the other hand, there is information stored as historical contingency and
memory. It is this latter kind that explains the thermodynamic functionality of
Maxwell's Demon, as it uses stored information about the thermal fluctuations
to convert them to useful work~\cite{Horo2010}. This recognition handily
resolves Maxwell's Second Law paradox. This information dichotomy was recently
laid bare by mapping Szilard's single-molecule engine to chaotic dynamical
system; a mapping so simple that all questions can be analytically addressed
\cite{Boyd14b}. The role of both informative measurement and its use, when
stored, for control illustrates the complementary role and functional
consequences of both kinds of information in an adaptive system.

In this way, the now-familiar physical setting of Maxwell's paradox highlights
how the distinction between \emph{information-as-surprise} and \emph{stored
actionable-information} motivated Ashby's emphasizing requisite
variety in adaptation. Detecting environmental fluctuations and acting on their structure (such as temporal correlations) are critical to the Demon's functioning. Appealing to new results
in nonlinear dynamics and nonequilibrium thermodynamics, the distinction
similarly motivates our re-addressing this central concern in cybernetics, so
basic to the operation of adaptive systems, but in a fully thermodynamic
setting: What requisite variety (range of historical contexts) must an adaptive
agent recognize in its environment to realize thermodynamic benefits? 

In the following, we first give an overview of our contributions (Sec.  \ref{sec:Synopsis}). We mention how Ashby's law of requisite variety is faithfully reflected in the behavior of { \it information engines}---autonomous versions of Maxwell's Demon. This close connection follows from the bounds set by the Second Law of Thermodynamics for information processing \cite{Mand012a, Boyd15a, Boyd16c}. Important for engineered and biological implementations, we note that these bounds, and so those specified by Landauer's Principle \cite{Land61a, Benn82}, are not generally achievable. The subsequent sections form the technical components of the development, key to which is representing an information reservoir in terms of the outcomes of a hidden Markov process.

Section \ref{sec:RequisiteRatchets} considers (i) the meaning of memory for the input processes of information engines and for the engines themselves, (ii) their energetics, and (iii) the role of memory in information thermodynamics more generally~\cite{Deff2013, Bara2014a}. It is the thermodynamics of memory that establishes the correspondence between Ashby's law and the behavior of information engines. Section \ref{sec:Achievability} addresses the limits on information engines achieving the informational Second Law bounds. We see that the bounds are not saturated even by optimal, finite-state engines. We also mention the curious case of infinite-memory information engines that can achieve and then go beyond these bounds, essentially by leveraging their internal infinite ``negentropy" to generate work~\cite{Bril51a}. These results bear directly on the description of Maxwell's original demon and, more contemporarily, stochastic universal Turing machines built out of information engines. Finally, we conclude with a summary of our results and their implications for biological physics and engineering.

\section{Synopsis: Memory and Thermodynamics}
\label{sec:Synopsis}

Szilard's Engine and related Maxwellian Demons are instances of thermal agents
processing environmental information in order to convert thermal energy into
work. Turning disordered thermal energy into work (ordered energy) was long
thought to violate the Second Law of Thermodynamics \cite{Benn87a}. However,
the past century resolved the apparent violation by recognizing that
information processing has unavoidable energy costs. Rolf Landauer was one of
the first to set bounds on information processing---specifically, erasing a
bit---such that the work production over a thermodynamic cycle cannot be
positive, satisfying the Second Law of thermodynamics \cite{Land61a, Benn82}.

However, if the Demon accesses an information reservoir in its environment, it can use the reservoir's statistics as a resource to convert thermal energy into work. This view of a Demon taking advantage of a structured environment connects back to cybernetics. Just as Ashby asked how a controller's variety should match that of its inputs, we ask how the Demon's internal structure should match the structure of an input process, which characterizes the information reservoir, in order to generate work. In contrast to cybernetics, though, we consider the variety inherent in ``information ratchets'' viewed as thermodynamic systems and, by implication, the variety they can detect and then leverage in their environments.

An information ratchet is an explicit construction of an autonomous Maxwellian Demon that uses an input symbol sequence to turn thermal energy into work energy \cite{Mand012a,Lu14a}. The ratchet steadily transduces the input symbols into an output sequence, processing the input information into an output while effecting thermodynamic transformations---implementing a physically embedded, real-time computation. This is accomplished by driving the ratchet along the input symbol sequence unidirectionally, so that the ratchet (with states in set $\mathcal{X}$) interacts once with each symbol (with values in alphabet $\mathcal{Y}$). During the interaction, the ratchet and current symbol make a joint transition from $x \otimes y \in \mathcal{X} \otimes \mathcal{Y}$ to $x' \otimes y'$ with probability \cite{Boyd15a}:
\begin{align*}
& M_{x \otimes y \rightarrow x' \otimes y'} \\
  & \quad\quad = \Pr(X_{N+1}=x',Y'_N=y'|X_N=x,Y_N=y)
  ~.
\end{align*} 
$M$ is a detailed-balanced Markov chain. (The requirement of detailed balance comes from the fact that the ratchets correspond to thermal physical systems performing computation and, thereby, must satisfy this condition in the absence of external, nonconservative forces, which we assume to be the case.) The transition matrix $M$ determines the energetics as well as the ratchet's information processing capacity.

\begin{figure}[tbp]
\centering
\includegraphics[trim = 0 0in 0 0in, width=\columnwidth]{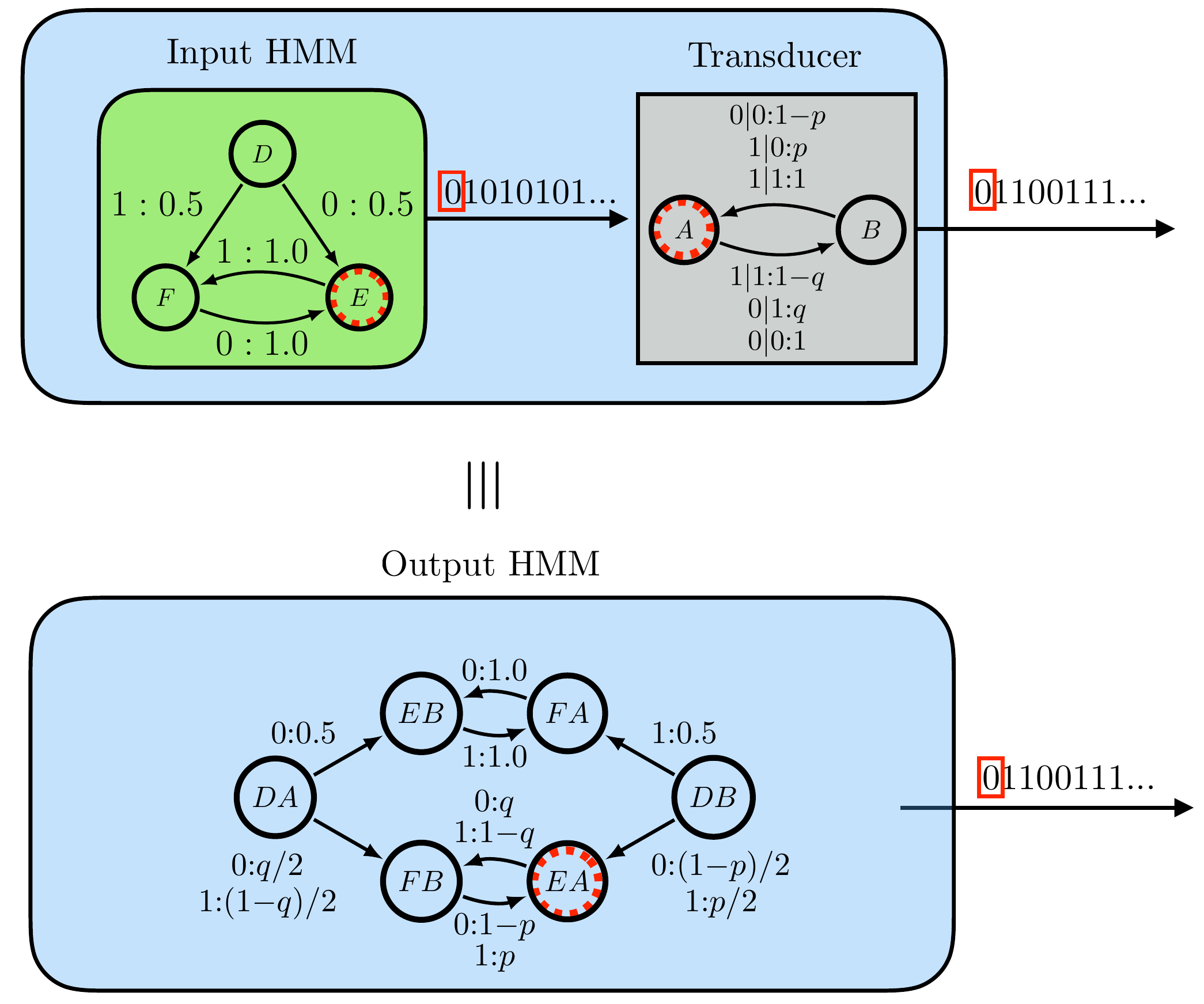}
\caption{Computational mechanics view of an information ratchet: The input
  signal (environment) is described by a hidden Markov model (HMM) that
  generates the input symbol sequence. The ratchet itself acts as a transducer,
  using its internal states or memory to map input symbols to output symbols.
  The resulting output sequence is described by an HMM that results from
  composing the transducer with the input HMM.  The current internal state of the input HMM, transducer, and output HMM are each highlighted by a dashed red circle.  These are the states achieved after the last output symbol (highlighted by a red box) of each machine.  We see that the internal state of the output HMM is the direct product of the internal state of the transducer and the input HMM.}
\label{fig:CompMechPic}
\end{figure}

Recently, we introduced a general computational mechanics \cite{Crut12a,Barn13a}
framework for analyzing thermodynamic devices that transduce an input process
into an output process \cite{Boyd15a, Barn13a}. Figure \ref{fig:CompMechPic}
depicts the relative roles of the input process specified by a finite-state
hidden Markov model (HMM), the ratchet as transducer operating on the input
process, and the resulting output process, also given by an HMM.

The tools of computational mechanics were developed to quantitatively analyze
how a ratchet's structure should match that of its input for maximum efficacy,
since they use a consistent notion of structure for general processes and
transformations. In particular, using them we recently established a general
information processing Second Law (IPSL) for thermodynamically embedded
information processing by finite ratchets that bounds the work $\langle W
\rangle$ in terms of the difference in entropy rates of the input and output
processes, $\hmu$ and $\hmu^\prime$, respectively \cite{Boyd15a}:
\begin{align}
\langle W \rangle \leq k_B T \ln 2 \left( \hmu^\prime - \hmu \right)
  ~.
\label{eq:EntropyRateSecondLaw}
\end{align}
(Definitions are given shortly in Sec. \ref{sec:RequisiteRatchets}.) Employing entropy rates---the Shannon entropy rate of the symbol sequence or, equivalently here, the Kolmogorov-Sinai entropy of its generative dynamical system---the bound accounts for all temporal correlations in the input and output processes as well as the single-symbol biases.
While this bound appears similar to that $\langle W \rangle \leq \langle I
\rangle-\Delta F$ ~\cite{Saga2010} on work production in a system with feedback
control, $\langle I \rangle$ quantifies correlations between the controller and
environment rather than temporal correlations induced in the environment.

Two uses of Eq. (\ref{eq:EntropyRateSecondLaw})'s IPSL suggest themselves. 
First, it sets an informational upper bound on the maximum average work production $\langle W \rangle$ per thermodynamic cycle. 
Here, $W$ is the flow of work from the ratchet to an external driver. 
Second, and complementarily, it places an energetic lower bound on the minimal work $\langle W_d \rangle $ required to drive a given amount ($\Delta \hmu$) of computation forward. 
Here, $W_d=-W$ is the flow of work from the driver into the ratchet. In this second use, the IPSL is a substantial extension of Landauer's Principle.
The latter says that erasing a bit of information requires a minimum energy expenditure of $\kB T \ln 2$ while the IPSL applies to any kind of computational processing that transforms an input process to an output process, not simply erasure. 
The first use appears, in
this light, as a rough converse to Landauer's limit: There is a potential
thermodynamic benefit of ``destroying variety'' in the form of
work~\cite{Land61a, Benn82}. 

Practically, computational mechanics gives a means to partition the ratchet and
input process into different cases: memoryful and memoryless. Whether or not
the input process or ratchet have memory substantially changes the bound on
work production. And so, we can examine how environmental and Demon varieties
interact. For example, in the case in which temporal correlations (varieties)
vanish, the difference between the single-symbol entropy of the input ($\H_1$)
and that of the output ($\H_1^\prime$) gives an analogous bound~\cite{Merh15a}:
\begin{align}
\langle W \rangle \leq k_B T \ln 2 \left( \H_1^\prime - \H_1 \right)
  ~,
\label{eq:SingleSymSecondLaw}
\end{align}
Using the single-symbol approximation $\H_1$ of the true entropy rate $\hmu$
can be quite convenient since $H_1$ is much easier to calculate than $\hmu$, as
the latter requires asymptotic (long-range) sequence statistics. (Again,
definitions are given shortly in Sec. \ref{sec:RequisiteRatchets}.) Likely,
this is why the $\H_1$-bound has appeared frequently to describe ratchet
information processing~\cite{Mand012a,Mand2013,Bara2013,Bara2014b,Bara2014a}.
Also, Eq. (\ref{eq:SingleSymSecondLaw}) is a rather direct generalization of
the Landauer limit, since the input entropy $\H_1 = 1$ bit and the output
$\H_1^\prime = 0$ bits saturate the bound on the work required to drive erasing
a binary symbol. However, a key difference is that Eq.
(\ref{eq:EntropyRateSecondLaw})'s entropy rates are dynamical invariants;
unchanged by smooth transformations \cite{Kolm59,Sina59}. The single-symbol
Shannon entropies are not dynamical invariants. In addition, the single-symbol
bound does not properly account for the temporal correlations in the input
process or those created by the ratchet in the output process and so leads to
several kinds of error in thermodynamic analysis. Let us explore these.

First, the average total temporal correlation in a process can be quantified
by the difference between the single-symbol entropy and the entropy rate, known
as a process' length-$1$ \emph{redundancy} \cite{Crut01a}:
\begin{align}
\H_1- \hmu \geq 0
  ~.
\label{eq:CorrelationMeasure}
\end{align}
This is the extent to which single-symbol entropy-rate estimates ($\H_1$)
exceed the actual per-symbol uncertainty ($\hmu$); and it is always
nonnegative. This measure describes a type of structure distinct from
statistical autocorrelations. Unless stated otherwise, going forward, the
informational temporal correlations quantified in Eq.
(\ref{eq:CorrelationMeasure}) are what we mean by \emph{correlations}.

How inputs or ratchets create or destroy these correlations determines
the relative strength and validity of the Eq. (\ref{eq:EntropyRateSecondLaw})
and Eq. (\ref{eq:SingleSymSecondLaw}) work bounds. These bounds, in turn,
suggest that memoryless ratchets are best for leveraging memoryless inputs and
memoryful ratchets are best for leveraging memoryful inputs and generating
work. However, it is not clear if and when the bounds are achievable. So, more
effort is required to establish this thermodynamic version of Ashby's Law of
Requisite Variety.

To address achievability, we turn to a general energetic framework for
calculating ratchet work production~\cite{Boyd16c}. There it was shown that
memoryful ratchets can leverage temporal correlations which memoryless ratchets
cannot. In short, memoryful ratchets are indeed best for leveraging memoryful
inputs. This gives an explicit violation of Eq. (\ref{eq:SingleSymSecondLaw}).
However, for memoryless ratchets both Eqs. (\ref{eq:SingleSymSecondLaw}) and
(\ref{eq:EntropyRateSecondLaw}) are valid bounds ~\cite{Merh15a}. We show, with
proof given in App. \ref{app:OptMemorylessInputs}, that memoryless ratchets are
the best among finite ratchets at leveraging statistical biases in memoryless
inputs to produce work. Notably, these ratchets do not achieve the derived
upper bounds on work production, demonstrating fundamental inefficiencies in
the information-to-work conversion in this class of an autonomous Maxwellian
Demon.

To approach the bounds described by Eqs. (\ref{eq:EntropyRateSecondLaw}) and
(\ref{eq:SingleSymSecondLaw}) it is necessary to go beyond the information
processing paradigm of a single finite-memory ratchet that interacts with a
single symbol at a time.  For instance, consider a ``swarm" of finely tuned
ratchets that work in a sequence, the output of one acting as the input of the
next, and each ratchet being optimized with respect to its own input. This
stepwise, sequential processing of the information reservoir is more efficient
than the single-ratchet paradigm and is able to approach the upper bounds on
information processing as the number of ratchets in the army grows. (This is
reminiscent of the higher efficiency of quasistatic thermodynamic processes
compared to finite-time, irreversible processes.) We reserve the detailed
analysis of this phenomenon for a later work since the framework for collective
thermodynamics is less developed than the single-ratchet setting we focus on
here.

While the IPSL and related bounds on work are suggestive of how the structure
of the input matches the output, the fact that they are unachievable for single
information ratchets means we must reach further to solidify the relationship
between input statistics and ratchet thermodynamics. Exact calculations here
for the work production verify the intuition that the memory of an optimal
ratchet must match the memory of the input. This leads to a variation on
Ashby's Law of Requisite Variety: ``memory leverages memory''.

In this way, the transducer framework for information ratchets gives insight
into how adaptive agents leverage structure. Its importance extends far beyond,
however, to general computation. On the one hand, transducers describe mappings
from input sequences to distributions over output sequences \cite{Barn13a,
Broo89a} and do so in real time. Turing machines, on the other, map individual
input sequences to individual output sequences with no particular reference to
physical time. In this sense, Turing machines are a subclass of transducers,
emphasizing that transducers are a general model for physical computation and
information processing. However, to do universal computation, as properly
configured Turing machines can, requires infinitely many states \cite{Broo89a}.
And, this suggests examining the thermodynamics of infinite-memory ratchets. 

It turns out that infinite ratchets with states having finite energy
differences are pathological in that they violate both the IPSL and its
single-symbol sister bounds on work
production---Eqs.~(\ref{eq:EntropyRateSecondLaw}) and
(\ref{eq:SingleSymSecondLaw}), respectively. The proof of Eq.
(\ref{eq:SingleSymSecondLaw}) assumes a stationary distribution over the
ratchet state and input symbol. This need not exist for infinite ratchets
\cite{Boyd15a}. In this case structure in the ratchet's memory, rather than
structure in the information reservoir, can be used as an additional
thermodynamic resource to produce work. And, this means that a framework for
general computation requires more detailed analysis to set bounds on work
production that account for the ratchet's memory. While we leave this for
upcoming work, it does call into question any discussion of the thermodynamics
of universal computation.

With this overview laid out, with the goals and strategy stated, we now are
ready to delve into memory's role in information-engine thermodynamics and the
achievability of the IPSL and its related bounds.

\begin{figure*}[tbp]
\centering
\includegraphics[width=2\columnwidth]{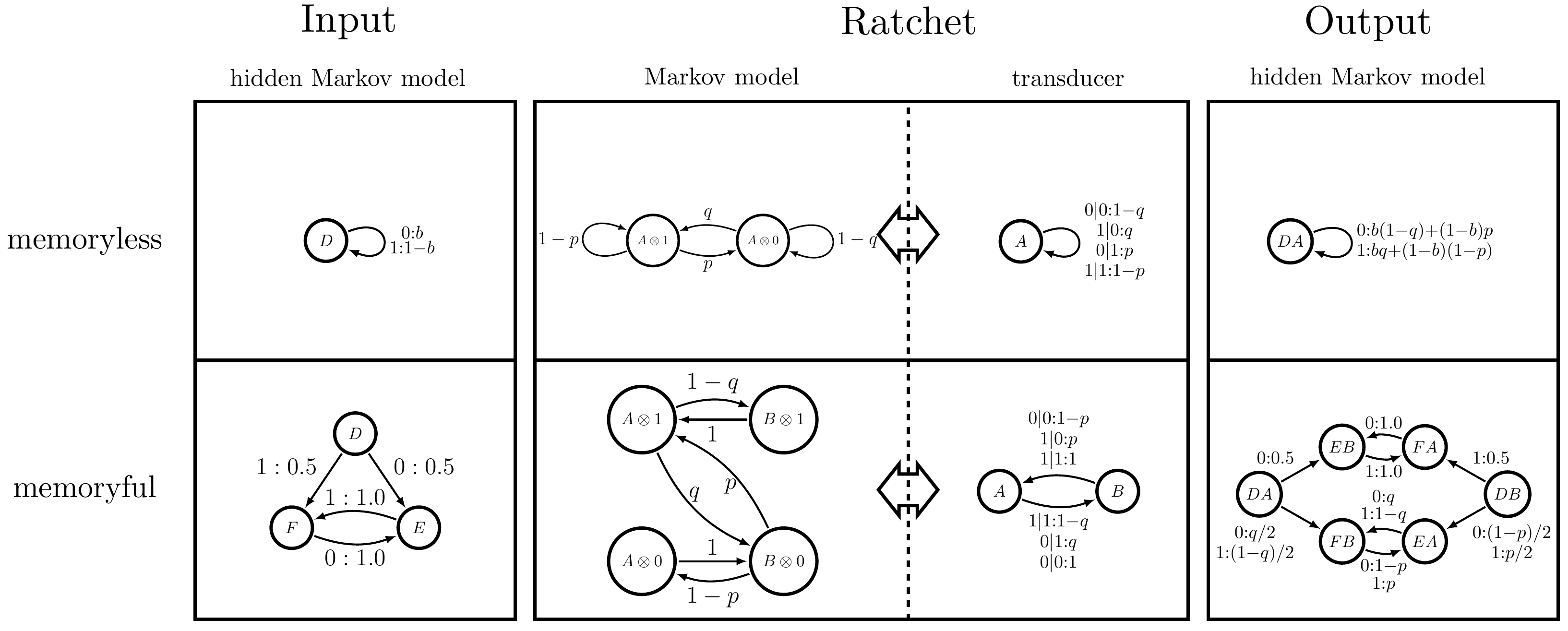}
\caption{Ratchets and input and output signals can be either memoryful or
	memoryless. For the input or output signal to be memoryless, the generating
	(minimal) HMM must have more than one internal state. The action of a
	ratchet can be represented in two different ways: either by a detailed
	Markov model involving the joint state space of the ratchet and an input
	symbol or by a symbol-labeled Markov dynamic on the ratchet's state space.
	We call the latter the \emph{transducer representation}~\cite{Barn13a}.
	Similar to the input and output signals, if the (minimal) transducer has
	more than one internal state, then the ratchet is memoryful.
  }
\label{fig:MemoryDifferences}
\end{figure*}

\section{Requisite Ratchets}
\label{sec:RequisiteRatchets}

To explore how a ratchet's structure ``matches'' (or not) that of an
environmental signal requires quantifying what is meant by structure. In terms
of their structure, both ratchets and environmental inputs can be either
memoryless or memoryful and this distinction delineates a ratchet's
thermodynamic functioning via the IPSL. This section introduces what we mean by
the distinction, describes how it affects identifying temporal correlations,
and shows how it determines bounds on work production and functionality. The
results, though, can be concisely summarized. Figure
\ref{fig:MemoryDifferences} presents a tableau of memoryless and memoryful
ratchets and inputs in terms of example HMM state-transition diagrams.
Figure~\ref{tab:BoundTable} then summarizes IPSL bounds for the possible
cases.  

\subsection{Process memory}

The amount of memory in the input or output processes is determined by the
number of states in the minimal representative dynamics that generates the
associated sequence probability distributions. While there are many ways to
generate a process, HMMs are a particularly useful representation of generating
mechanisms. For example, they describe a broader class of processes than
finite-order Markov models, since they can generate infinite Markov-order
processes using only a finite number of hidden states \cite{Crut01a}.

Here, we use the Mealy representation of HMMs~\cite{Rabi86a, Rabi89a, Elli95a,
Ephr02a}, which consists of a set $\CausalState$ of internal states and an
alphabet $\ProcessAlphabet$ of symbols that are emitted. As with a Markov
chain, transitions between hidden states in $\mathcal{S}$ are made according to
conditional probabilities. However, the generated symbols in $\mathcal{Y}$ are
emitted during transitions between hidden states, rather than when entering
states \cite{Shal98a}. The Mealy HMM dynamic is specified by a set of
symbol-labeled transition matrices:
\begin{align*}
T^{(y_N)}_{s_N \rightarrow s_{N+1}}=\Pr(Y_N=y_N,S_{N+1}=s_{N+1}|S_N=s_N)
  ~,
\end{align*}
which give the joint probability of emitting $y_N$ and transitioning to hidden
state $s_{N+1}$ given that the current hidden state is $s_N$. For the special
class of \emph{unifilar} HMMs the current hidden state $s$ and emitted symbol
$y$ uniquely determine the next hidden state $s'(s, y)$. Helpfully, for
unifilar HMMs the generated process' entropy rate $\hmu$ is exactly given by
the state-averaged uncertainty in the emitted symbols given the current state
\cite{Crut01a}:
\begin{align*}
\hmu & = \lim_{N \rightarrow \infty} \H[Y_N|Y_{0:N}] \\
     & = \lim_{N \rightarrow \infty} \H[Y_N|S_N] \\
     & = \sum_{s \in \mathcal{S}}
		 \pi_s \lim_{N \rightarrow \infty} \H[Y_N|S_N=s] \\
	 & = -\sum_{s \in \mathcal{S}}
		 \pi_s \sum_{y\in \mathcal{Y}}
	T_{s \rightarrow s'(s, y)}^{(y)}
	 \log_2
	T_{s \rightarrow s' (s, y)}^{(y)}
  ~,
\end{align*}
where $\pi_s$ is the steady-state distribution over the hidden states. A
process' \emph{\eM} is its minimal unifilar HMM generator, where minimality is
determined by having the smallest internal-state Shannon entropy \cite{Crut12a}:
\begin{align*}
\lim_{N \rightarrow \infty}\H [S_N] &= -\lim_{N \rightarrow \infty}\sum_{s \in \mathcal{S}} \Pr(S_N=s) \log_2{\Pr(S_N=s)}
\\ &  = -\sum_{s \in \mathcal{S}} \pi_s \log_2{\pi_s}
\\ & \equiv \Cmu
  ~. 
\end{align*}
where in the last line we defined the process' \emph{statistical complexity}
$\Cmu$. Since $\hmu$ gives an exact expression for process entropy rate and
$\Cmu$ a unique definition of process memory \cite{Crut92c}, throughout we
represent processes by their \eMs. An \eM's internal states are called
\emph{causal states}.

Broadly, the memory of an \eM\ refers to its hidden states. As shown in Fig.
\ref{fig:MemoryDifferences}, memoryless input processes have \eMs\ with a
single state: $|\mathcal{S}|=1$. The sequence distributions for such processes
are given by a product of single-symbol marginal distributions.  For a
stationary process, the single-symbol marginal entropy $\H_1$ is the same for
every symbol:
\begin{align}
\H_1 \equiv \H[Y_N]  \text{ for all } N \in \mathbb{N}.
\label{eq:SingleSymbol}
\end{align}
For memoryless processes, the entropy rate is the same as the single-symbol entropy:
\begin{align*}
h_\mu & = \lim_{N \rightarrow \infty} \H[Y_N|Y_{0:N}]
\\&=\lim_{N \rightarrow \infty} \H[Y_N]
\\ & = \H_1
  ~.
\end{align*}
This means that their difference vanishes:
\begin{align}
\label{eq:MemorylessSignal}
\H_1 - \hmu = 0
  ~,
\end{align}
and, thus, there are no temporal correlations in the symbol string, because
$\H_1-h_\mu$ quantifies the informational correlation of individual input
symbols with past inputs:
\begin{align}
\H_1-h_\mu & = \lim_{N \rightarrow \infty}
  ( \H[Y_N] - \ H[Y_N|Y_{0:N}]) \nonumber \\
  & = \lim_{N \rightarrow \infty} \I[Y_N : Y_{0:N}]
  ~,
\label{eq:CorrelationMeasure2}
\end{align}
where $\I[W:Z]$ is the mutual information of random variables $W$ and $Z$
\cite{Cove06a}.

For memoryful input processes, as shown in Fig.
\ref{fig:MemoryDifferences}, there are multiple causal states for the \eM:
$|\mathcal{S}| >1$. In other words, sequence probabilities cannot be broken
into a product of marginals. And so, in general, we have:
\begin{align*}
\H_1 > h_\mu
  ~.
\end{align*}
Thus, there are temporal correlations in the input process:
\begin{align}
\H_1 - h_\mu > 0
  ~.
\end{align}
This means that individual symbols of the input sequence share information with
past inputs. In the maximally correlated case, every symbol is exactly
predictable from its past. As a result the entropy rate vanishes and the
temporal correlation measure in Eq. (\ref{eq:CorrelationMeasure2}) is equal to
the single-symbol entropy.

To summarize, memoryless input signals have a single causal state and, thus, do
not exhibit temporal correlations, since they have no way to store information
from the past. Meanwhile, memoryful inputs have multiple hidden states that are
used to transmit information from the past to the present and so express
temporal correlations.

\subsection{Ratchet memory}

From the perspective of information processing, the ratchet is a transducer
that interacts with each symbol in the input sequence in turn, converting it
into a output symbol stored in the output sequence~\cite{Barn13a,Boyd15a}. The
ratchet is a form of communication channel \cite{Cove06a}. One that is
determined by a detailed-balanced Markov dynamic:
\begin{align*}
& M_{x_N \otimes y_N \rightarrow x_{N+1} \otimes y'_N} 
\\ & = \Pr(Y'_N=y'_N,S_{N+1}=s_{N+1}|S_N=s_N, Y_N = y_N)
 ~
\end{align*}
over the ratchet's state space $\mathcal{X}$ and a symbol alphabet
$\mathcal{Y}$. This is the probability that the ratchet ends in state $x_{N+1}$
and writes a symbol $y'_N$ to the output sequence, given that the input symbol
was $y_N$ and the ratchet's state was $x_N$ before the symbol-state interaction
interval.

The Markovian dynamic describes the behavior of the joint event (ratchet-state
$\otimes$ symbol-value) during the interaction transition and leads to the
transducer representation of the ratchet's functionality, illustrated in
Fig.~\ref{fig:MemoryDifferences}. As we use the terms, the \emph{ratchet} refers
to the physical device implementing the Markovian dynamic, whereas
\emph{transducer} refers to the computational mechanics state-transition
machine (\eT) that captures its information-theoretic functionalities in a
compact way ~\cite{Barn13a}. The form of the transducer is:
\begin{align}
M_{x_N \rightarrow x_{N+1}}^{(y'_N|y_N)}
  = M_{x_N \otimes y_N \rightarrow x_{N+1} \otimes y'_N}
  ~.
\end{align}
The distinction between the Markov dynamic and the transducer representation is
best illustrated graphically, as in the second column of
Fig.~\ref{fig:MemoryDifferences}.

The definition of a ratchet's memory involves its $\epsilon$-transducer
representation. In other words, memory is related to the size of the ratchet's
causal state space $|\mathcal{X}|$ in its $\epsilon$-transducer representation.
(The very definition of \eMs\ and \eTs\ entails that they have the minimal set
of states for a given input, output, or input-output process.) As seen in the
top middle of Fig. \ref{fig:MemoryDifferences}, memoryless ratchets have only a
single internal (hidden) state: $|\mathcal{X}|=1$. Thus, the ratchet behaves as
a memoryless channel from input to output~\cite{Cove06a}. And, in this, it
reduces temporal correlations in the input signal:
\begin{align}
\label{eq:MemorylessRatchet}
\H_1^\prime - \hmu^\prime \leq \H_1 - \hmu
  ~,
\end{align}
according to the channel coding theorem \cite{Cove06a}. That is, the change in
single-symbol entropy is a lower bound for the change in entropy
rates~\cite{Merh15a}. In contrast, a memoryful ratchet has more than one state,
$|\mathcal{X}|>1$, and behaves as a memoryful channel~\cite{Barn13a}; bottom
right of Fig. \ref{fig:MemoryDifferences}.

How the ratchet transduces the
current input to the current output depends on in which state it is.  As a
result, the ratchet can create correlations in the output such that, regardless
the input process:
\begin{align}
\label{eq:MemoryfulRatchet}
\H_1^\prime - \hmu^\prime \geq 0
  ~.
\end{align}
Several explicit constructions of the output process based on given input and ratchet are shown in the last column of Fig.~\ref{fig:MemoryDifferences}.

\begin{figure*}[tbp]
\centering
\includegraphics[width=2\columnwidth]{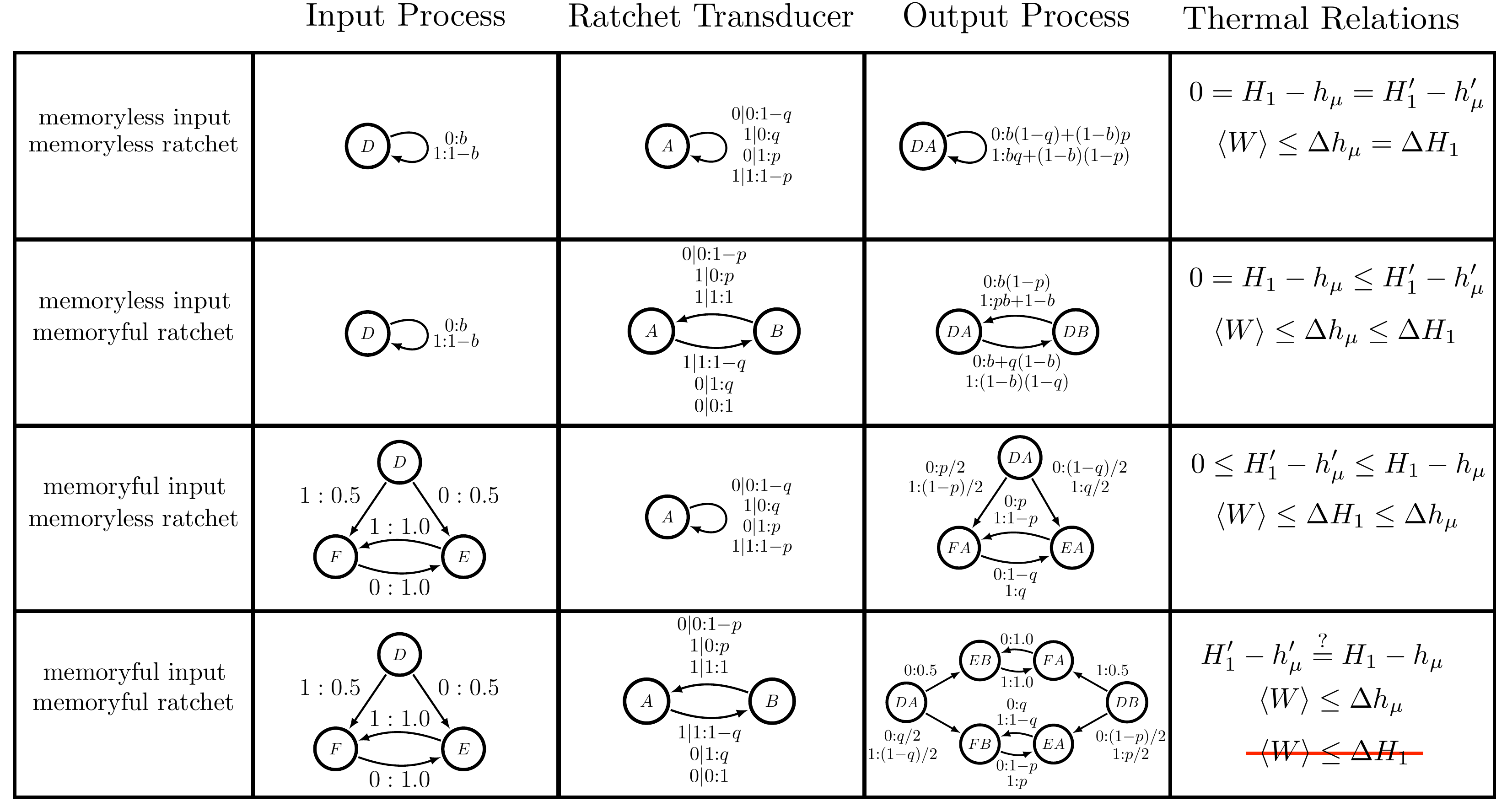}
\caption{The informational (IPSL) bounds on work that use $\Delta \hmu$ or
	$\Delta \H(1)$ depend critically on input signal and ratchet memory. In all
	finite memory cases, $\Delta \hmu$ is a valid bound on $\frac{\langle W
	\rangle}{k_B T \ln2 }$, but the same is not true of $\Delta \H_1$, as
	indicated in the far right column on thermal relations.  The bounds shown
	in column have $k_B T \ln 2$ set to unity, so that the relations can be
	shown in compact form. If the ratchet is memoryless, then $\Delta \H_1$ is
	a valid and stronger bound than $\Delta \hmu$, because these channels
	decrease the temporal temporal correlations in transducing input to
	output.  For a memoryless input with memoryful ratchet, $\Delta \H_1$ is
	still a valid bound, but it is a weaker bound than $\Delta \hmu$, because
	memoryful ratchet can create temporal correlations in the output. However,
	in the case where both input and output are memoryful, the $\Delta \H_1$
	bound is invalid and can be violated by systems which turns temporal
	correlations into work by having the memory of the ratchet synchronize to
	the memory the input.
	}
\label{tab:BoundTable}
\end{figure*}

\subsection{Thermodynamics of memory}

This section considers the role of memory in the thermodynamic efficacy of
information engines. In particular, we consider the average work production per
cycle $\langle W\rangle$. The role can be explored in two complementary ways:
either following the IPSL and related bounds,
Eqs.~(\ref{eq:EntropyRateSecondLaw}) and (\ref{eq:SingleSymSecondLaw}) or from
the exact expression of $\langle W \rangle$.

\subsubsection{Information-processing Second Law bounds}

The thermodynamics of memory is summarized in Fig. ~\ref{tab:BoundTable}'s
table, where each row considers a different combination of input process and
ratchet. This section addresses each cell in the table individually.

Consider the case of memoryless input and a memoryless ratchet. In
Eq.~(\ref{eq:MemorylessRatchet}), we saw that the temporal correlations in the
input signal cannot be increased by such ratchets. Since the input signal is
memoryless, the output signal must also be memoryless. For memoryless signals,
however, we saw via Eq.~(\ref{eq:MemorylessSignal}) that the entropy rate
$\hmu$ is the same as the single-symbol entropy $\H_1$. We conclude that the
single-symbol entropy input-to-output difference is the same as the entropy-rate difference:
\begin{align*}
	\H_1^\prime - \hmu^\prime = \H_1 - \hmu
  	~.
\end{align*}
As a result both Eqs.~(\ref{eq:EntropyRateSecondLaw}) and
(\ref{eq:SingleSymSecondLaw}) give the same bound on the the average rate of
work production:
\begin{align}
	\langle W \rangle & \leq k_B T \ln2 ~\Delta \H_1 \\
                  & = k_B T \ln2 ~\Delta \hmu
  	~,
\end{align}
This is noted at the right column in the table's first row.
	 
Consider now the case of memoryful input with, again, a memoryless ratchet.  A
memoryful input contains temporal correlations that are decreased by the
memoryless ratchet, from Eq.~(\ref{eq:MemorylessRatchet}). The same equation
implies that the single-symbol entropy difference is an upper bound on the
entropy-rate difference. As a result, Eq.~(\ref{eq:SingleSymSecondLaw})
provides a quantitatively tighter bound on the work production compared to
the IPSL of Eq.~(\ref{eq:EntropyRateSecondLaw})~\cite{Merh15a}:
\begin{align*}
\langle W \rangle & \leq k_B T \ln2 ~\Delta \H_1 \\
                  & \leq k_B T \ln2 ~\Delta \hmu
  ~,
\end{align*}	

These observations suggest that memoryless ratchets cannot leverage 
temporal correlations, since the stricter bound (single symbol) on work 
production stays fixed as we hold the single-symbol entropy fixed but vary 
the temporal correlations in the input. It appears that to leverage temporal 
correlations, one must use a memoryful ratchet.
	
We now address the case of memoryful ratchets. First, consider the case of
memoryless inputs (no temporal correlations: $\hmu = \H_1$). From
Eq.~(\ref{eq:MemoryfulRatchet}), we know that memoryful ratchets can create
correlations in the output. In other words, the output signal is generally
memoryful, implying $\H'_1 - \hmu' \geq 0$. As a result, $\Delta \hmu$ is a
stricter bound than $\Delta \H_1$:
\begin{align*}
\langle W \rangle & \leq k_B T \ln2 ~\Delta \hmu \\
                  & \leq k_B T \ln2 ~\Delta \H_1
  ~,
\end{align*}
as seen in Table \ref{tab:BoundTable}'s second row. We explored this in some
detail previously \cite{Boyd15a}. By calculating $\Delta \hmu$, we found a
novel type of functionality in which the ratchet used stored work energy to
increase temporal correlations in the input while simultaneously
\emph{increasing} the single-symbol uncertainty. 
The above relations also imply that memoryless
ratchets may be best suited for leveraging memoryless input processes, since
the bounds on work production for memoryless ratchets are higher than the
bounds for memoryful ratchets.

Consider now a memoryful input driving a memoryful ratchet. In this
case, memory in the ratchet is useful for work production. In a companion work
\cite{Boyd16c} we consider a maximally correlated, period-$2$ input process,
that has no single-symbol negentropy to leverage ($\H_1=1$ bit of information),
but that has maximal temporal correlations ($\H_1 - \hmu=1$ bit). Notably, the
single-symbol bound indicates that no work can be produced, since $\Delta \H_1
\leq 0$ regardless of the output. Critically, though, the IPSL bound indicates
that work production is possible, since $\hmu^\prime - \hmu > 0$ as long as the
output has some uncertainty in each sequential symbol. Indeed, Ref.
\cite{Boyd16c} constructs a ratchet that produces positive work: $\langle W
\rangle = \kB T \frac{1-\delta}{e}$, where $\delta \in (0,1)$. Thus, the
single-symbol bound is violated, but the IPSL bound is satisfied, as shown in
Fig.~\ref{tab:BoundTable}'s last row.

The final case to consider, in fact, is left out of Fig. \ref{tab:BoundTable}:
infinite-memory ratchets. This is because infinite memory ratchets do not
necessarily have a steady state, so establishing the IPSL bound in Ref.
\cite{Boyd15a} does not hold. There are, as yet, no predictions for
infinite-memory ratchets based on the information measures of the input or
output processes.  However, this is an intriguing case. And so, we return to
the case of infinite ratchets in Sec.~\ref{sec:InfiniteMemoryRatchets}.

Stepping back, Fig. \ref{tab:BoundTable}'s table details a constructive
thermodynamic parallel to Ashby's Law of Requisite Variety: \emph{Memory can
leverage memory}. However, the bounds do not constitute existence proofs, since
it is not yet known if the specified bounds are achievable. Though, we
constructed an example of a temporally correlated process that is best
leveraged by memoryful ratchets, it is possible that there is an alternative
temporally correlated input process that is best leveraged by a memoryless
ratchet. Similarly, we see that the bounds on memoryless inputs are stricter
for memoryful ratchets than for memoryless ratchets. If these bounds are not
achievable, however, then this does not translate into a statement about the
ratchet's actual efficiency in producing work.

Before addressing this puzzle, we need to determine the work production.

\subsubsection{Exact work production}
\label{sec:Generalized Energetics}

An exact expression for the average work production rate was introduced in
 Ref. \cite{Boyd16c}:
\begin{align}
\langle W \rangle = k_B T
  \!\!\!  \sum_{ \substack{x,x' \in \mathcal{X} \\ y,y' \in \mathcal{Y} } }
  \!\! \pi_{x \otimes y}
  M_{x \otimes y \rightarrow x' \otimes y'}
  \ln \! \frac{M_{x' \otimes y' \rightarrow x \otimes y}}
  {M_{x \otimes y \rightarrow x' \otimes y'}},
\label{eq:Work}
\end{align}
where $\{\pi_{x \otimes y}\}$ is the steady-state joint probability
distribution of the ratchet and the input symbol before interaction.
Heuristically, the formula can be understood in the following way. At the
beginning of the interaction interval, the ratchet and the incoming bit have
probability $\pi_{x \otimes y}$ to be in state $x \otimes y$. Thus, the joint
system has the probability $\pi_{x \otimes y} M_{x \otimes y \rightarrow x'
\otimes y'} $ to make the transition  $x \otimes y \rightarrow x' \otimes y'$.
In each such transition, the amount of energy extracted from the reservoir is
given by the log-ratio  $\ln \left(M_{x' \otimes y' \rightarrow x \otimes y}/
M_{x \otimes y \rightarrow x' \otimes y'} \right)$. The right-hand side of
Eq.~(\ref{eq:Work}) therefore gives the average energy extracted from the heat
reservoir every thermodynamic cycle. From the First Law of Thermodynamics, this
must be the average work production by the ratchet, since the ratchet's energy
is fixed in the steady state.  Not only does the expression confirm our
physical law of requisite memory, it also expands our understanding of the
validity of IPSL-like bounds, as we see below.

Irrespective of the nature of the input, consider the case of memoryless
ratchets for which we have:
\begin{align*}
\pi_{x\otimes y} & = \lim_{N \rightarrow \infty} \Pr(X_N=x,Y_N=y) \\
                 & = \lim_{N \rightarrow \infty} \Pr(Y_N=y)
                 \\ & = \Pr(Y_N=y)
                 ~,
\end{align*}
simply the single-symbol probabilities of the input process. This follows since
there is only a single ratchet state $x$. Thus, from Eq. (\ref{eq:Work}), the
only dependence the work has on the input process is on the latter's
single-symbol distribution. In short, memoryless ratchets are insensitive to
correlations in the inputs. To leverage correlations beyond single symbols in
the input process it is necessary to add memory to the ratchet, as discussed in
the previous section and in our companion work~\cite{Boyd16c}.

Conversely, as App.~\ref{app:OptMemorylessInputs} establishes, if the input
process is memoryless, there is no energetic advantage of using finite
memoryful ratchets for binary input processes. For any finite memoryful ratchet
that extracts work using the input process, there exists a memoryless ratchet
that extracts at least as much work.

These two results confirm the intuition that to be thermodynamically optimal a
ratchet's memory must match that of the input: Memoryful ratchets best leverage
memoryful inputs and memoryless ratchets best leverage memoryless inputs.

\section{Achievability of Bounds}
\label{sec:Achievability}

The IPSL bound on average work production rate was derived based on the Second
Law of Thermodynamics applied to the joint evolution of the ratchet, the
input-output symbol sequence, and the heat reservoir. Since the Second Law is
merely an inequality, it does not guarantee that the bounds are actually
achievable, at least for the class of information engines considered here. In
point of fact, we saw that the bound cannot be saturated by memoryless
ratchets.  A somewhat opposite picture is presented by infinite-memory
ratchets.  And, understanding these is a necessity if we wish to build a
thermodynamics of general computation; that is, of physically embedded
universal Turing machines.  As we will show shortly, infinite-memory ratchets
can violate the IPSL bound since they can leverage the steady, indefinite
increase in their own entropy to reduce the entropy of the heat reservoir, in
addition to the contributions from an input signal. The following analyzes
these cases individually.

\subsection{Memoryless ratchets}
\label{sec:MemorylessRatchetEnergetics}

This section applies the work expression of Eq.~(\ref{eq:Work}) to find optimal
memoryless ratchets and then compares the optimal work production to the above
information thermodynamic bounds to determine their achievability.
Understanding the relationships between the memory of the ratchet and that of
the input process, as discussed above, deepens the interpretation of the
analysis. Since memoryless ratchets are insensitive to correlations, our
calculated work productions are not only the work productions for memoryless
inputs, but the work productions for all inputs with the same single-symbol
statistical biases.

A memoryless ratchet's memory consists of a single state. As a result, the
Markovian dynamic $M$ acts only on individual input symbols. Thus, the work for
any input process is a function only of its single-symbol distribution
$\pi_y=\Pr(Y_N=y)$ (given $M$):
\begin{align*}
\langle W \rangle = k_B T
  \sum_{y,y' \in \mathcal{Y}}
  \pi_y M_{y \rightarrow y'}
  \ln \frac{M_{y' \rightarrow y}}{M_{y \rightarrow y'}}
  ~.
\end{align*}

\begin{figure}[tbp]
\centering
\includegraphics[width=\columnwidth]{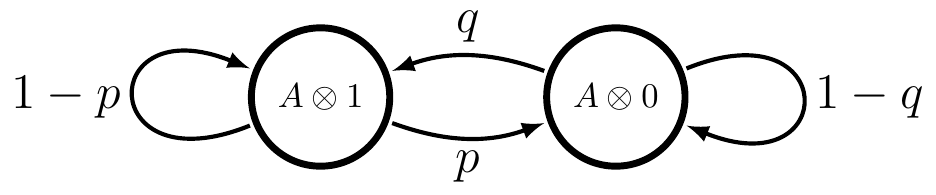}
\caption{All possible memoryless ratchets that operate on a binary input,
  parametrized by transition probabilities $p$ and $q$.
  }
\label{fig:Memoryless}
\end{figure}

Here, we discuss in detail the particular case of a memoryless ratchet driven
by binary inputs. The relevant class of transducers comprises all
two-state HMMs over the state space $\{A\} \otimes \{0,1\}$, where $A$ is the
sole state of the ratchet. Since the state space of the transducers is
two-dimensional, the Markovian dynamic $M$ is guaranteed to be detailed
balanced. Moreover, we can parametrize this class by two transition
probabilities $p$ and $q$, as shown in Fig. \ref{fig:Memoryless}. This, then,
allows us to optimize over $p$ and $q$ to maximize work production.

\begin{figure}[tbp]
\centering
\includegraphics[width=.9\columnwidth]{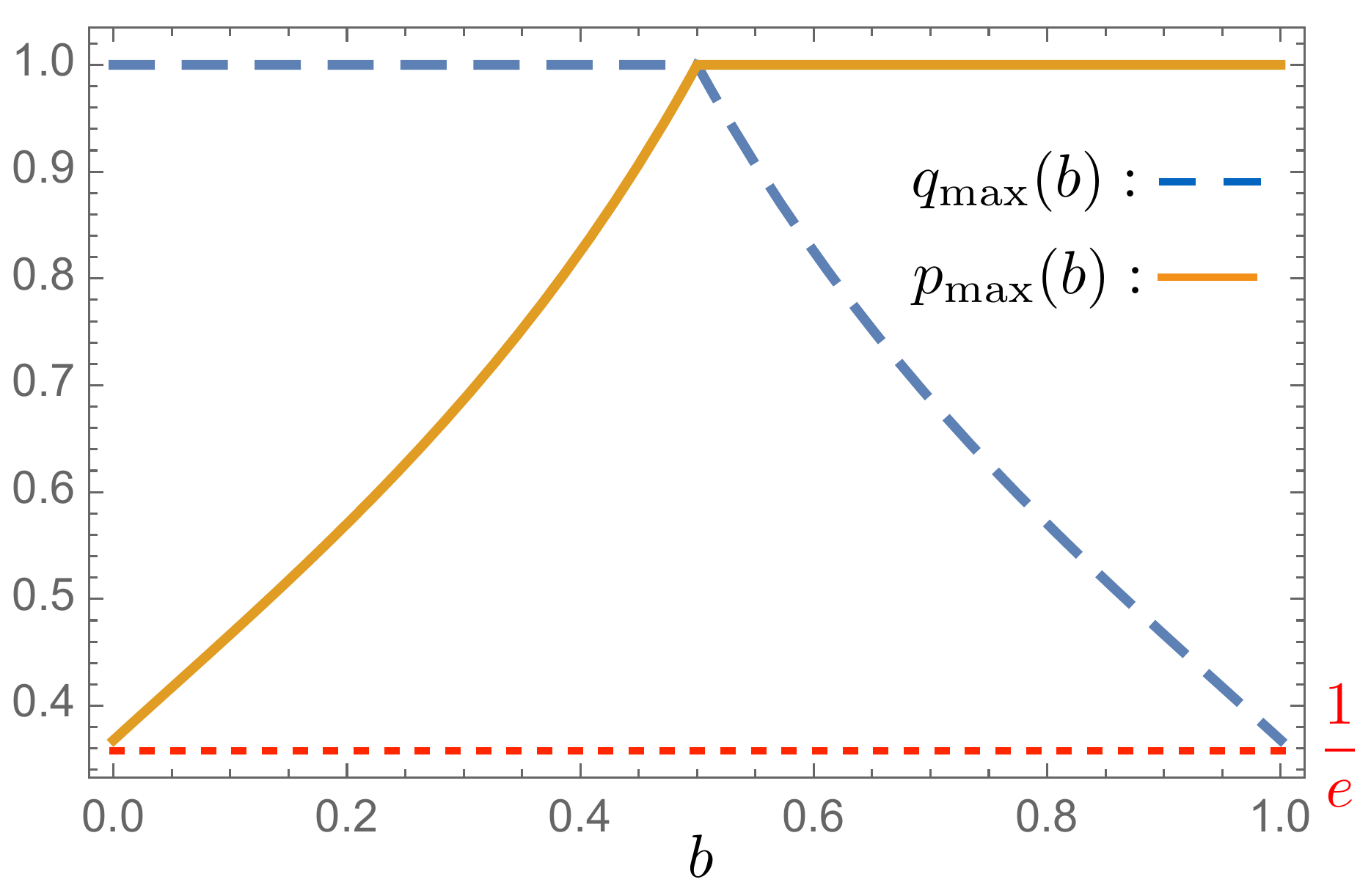}
\caption{Optimal ratchet parameters $p_\text{max}(b)$ (solid orange line) and
	$q_\text{max}(b)$ (dashed blue line) are mirror images about $b=1/2$. For
	$b<1/2$, we set $p_\text{max}(b)<1$ and $q_\text{max} =1$ so that the
	interaction transition $1 \rightarrow 0$ has a positive energy change
	$\Delta E_{1\rightarrow 0}= k_B T \ln (q/p)$ and, thus, absorbs heat from
	the thermal reservoir. The same reasoning applies to $b>1/2$, where
	$p_\text{max}(b)=1$ and $q_\text{max} <1$. In the unique case where the
	input is all $1$s, the most effective ratchet for generating work has
	$p_\text{max}=1/e$. Both functions realize a minimum value of $1/e$.
  }
\label{fig:MaxParameters}
\end{figure}

For the ratchet shown in Fig. \ref{fig:Memoryless} driven by a process with
single-symbol probabilities $\Pr(Y_N=0)=b$ and
$\Pr(Y_N=1)=1-b$, the average work done is a function of $b$, $p$, and $q$:
\begin{align}
\langle W \rangle(b,p,q) = k_B T (b - b') \ln\frac{p}{q}
  ~,
  \label{eq:WorkBinary}
\end{align}
where $b' = b'(b, p, q) = (1-q)b + (1-b)p$ is the probability of symbol $0$ in
the output:  $ \Pr(Y'_N=0)$.  The expression for $b'$ follows from the dynamic
depicted in Fig.~\ref{fig:Memoryless}, whereas Eq.~(\ref{eq:WorkBinary})
follows from the fact that work $\ln(p/q)$ is gained for each transformation $0
\rightarrow 1$. For a given input bias $b$, optimization of the ratchet's
transducer dynamic to produce maximal work yields ratchet parameters
$p_\text{max}(b)$ and $q_\text{max}(b)$:
\begin{align*}
\left[q_\text{max}(b),p_\text{max}(b)\right]
  = \begin{cases}
      [ \frac{1-b}{b \, \Omega(e(1-b)/b)},1] ,   & 1/2 \leq b \leq 1\\
      [ 1,\frac{b}{(1-b) \, \Omega(eb/(1-b))} ] , & 0 \leq b < 1/2
  \end{cases}
  ~,
\end{align*}
where the function $\Omega(\cdot)$ is defined implicitly as $\Omega(ze^z)=z$.
 Note that
there is a symmetry with respect to the simultaneous exchanges: $\{p
\leftrightarrow q, b \leftrightarrow 1-b\}$. Figure~\ref{fig:MaxParameters}
shows how the optimal parameters depend on input bias $\Pr(Y_N=0)=b$. Since the
ratchet is insensitive to temporal correlations, this behavior holds for all
input processes, temporally correlated or uncorrelated, where the probability
of a $0$ is $b$.

\begin{figure}[tbp]
\centering
\includegraphics[width=\columnwidth]{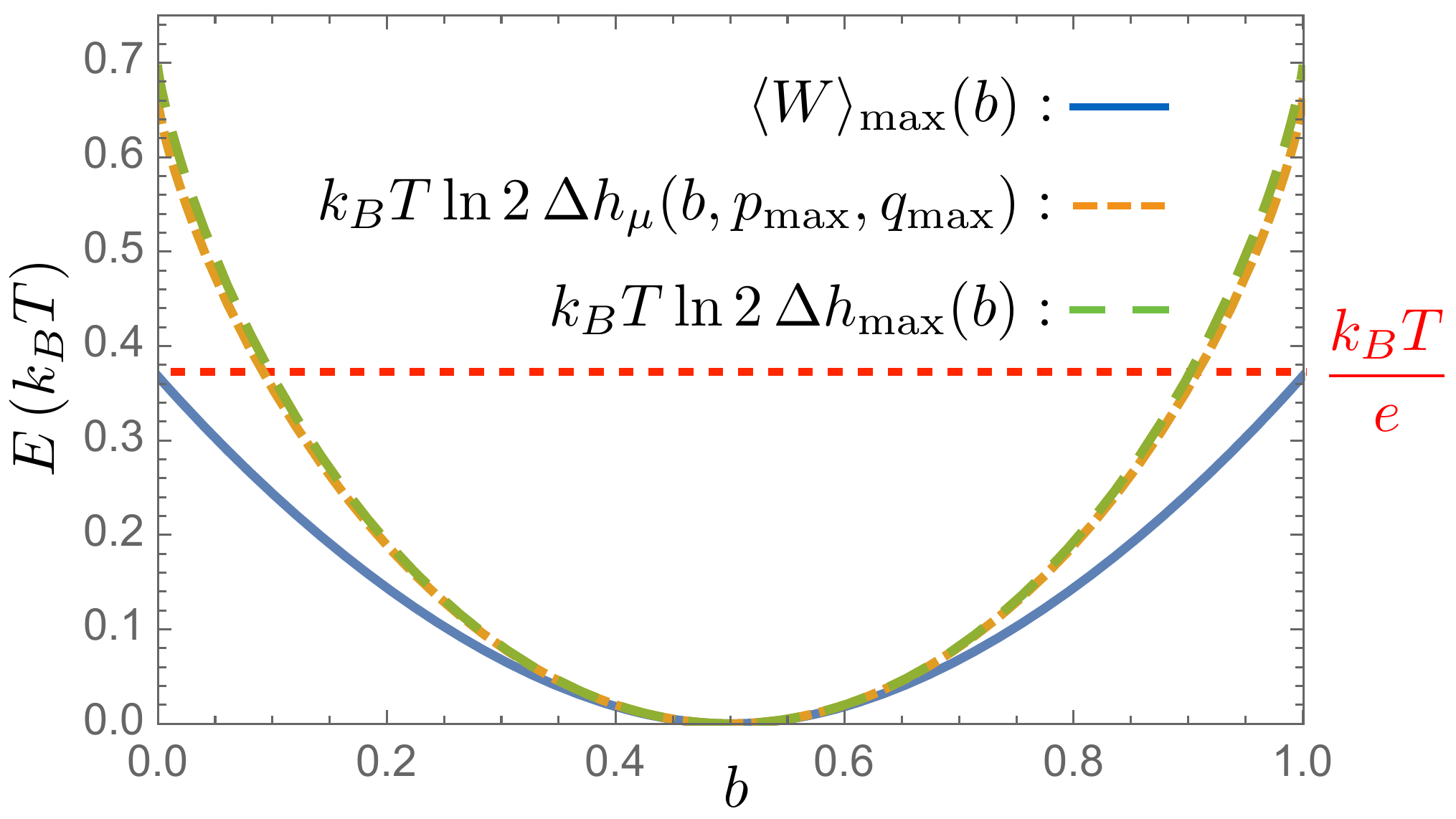}
\caption{Maximum work production $\langle W \rangle_\text{max}$ for any input
	bias $b$ is $\kB T/e$ (horizontal dashed line) and so ratchets do not
	achieve the IPSL upper bound $\langle W \rangle \leq k_B T \Delta
	h_\mu(b,p_\text{max},q_\text{max})$ that derives from pure informational
	properties of the input and output processes. Also, $\Delta
	h_\mu(b,p_\text{max},q_\text{max})$ itself is slightly less than the
	absolute maximum possible change in entropy $\Delta h_\text{max}(b)$ given
	an input bias $b$. This means that a memoryless ratchet does not leverage
	all of the single-symbol statistical order in the input.
  }
\label{fig:MemorylessWork}
\end{figure}

Substituting $q_\text{max}$ and $p_\text{max}$ into the expression for work
production, we find the maximum work production:
\begin{align*}
\langle W \rangle_\text{max}(b) =
  \langle W \rangle (b,p_\text{max}(b),q_\text{max}(b))
  ~,
\end{align*}
yielding the solid (blue) curve in Fig. \ref{fig:MemorylessWork}. The curve is
the maximum work production $\langle W \rangle_\text{max} (b)$ of a memoryless
ratchet for any input, temporally correlated or not, with bias $b$. 

To compare work production directly with the IPSL and related bounds,
Eqs.~(\ref{eq:EntropyRateSecondLaw}) and (\ref{eq:SingleSymSecondLaw}), we need
to calculate the changes in single-symbol entropy difference ($\Delta \H_1$) and
entropy-rate difference ($\Delta \hmu$). Reminding ourselves that the ratchet
is memoryless, these differences are the same if we assume the input to be
memoryless. We find:
\begin{align*}
\Delta \H_1 & = \Delta \hmu (b,p,q)  \\
            & = \H_\text{B}(b') - \H_\text{B}(b)
~, 
\end{align*}
with $\H_\text{B}(z) = \H(\{z, 1-z\})$ for $z \in [0,1]$, the binary entropy
function \cite{Cove06a}. We obtain the bounds for an optimal ratchet, for a
given input bias $b$, by substituting $p_\text{max}$ and $q_\text{max}$ for $p$
and $q$, respectively. We plot this optimal bound as the dashed line (orange)
in Fig. \ref{fig:MemorylessWork}. Even though we maximized over the memoryless
ratchet's parameters ($p$ and $q$), the output work $\langle W \rangle_\text{max}(b)$ falls far short of the
bounds set on it, as the solid (blue) curve lies below the dashed (orange)
curve except exactly at $b=1/2$, where there is zero work production. This
demonstrates that there are inherent inefficiencies in memoryless information
ratchets.

There is a second source of inefficiency for memoryless ratchets. The maximum
possible bound for the generated work comes from the case where there are no
statistical biases and no correlations left in the output sequence, so that the
output has maximum Shannon entropy. In this case we have $b' = 1/2$, the
maximal entropy change being:
\begin{align*}
\Delta h_\text{max}(b) = 1 - \H_B(b)
  ~.
\end{align*}
Figure \ref{fig:MemorylessWork} plots the corresponding bound as a dashed line
(green), showing that it lies above the actual change in entropy for an optimal
ratchet. Thus, not all of the order in the input sequence is being leveraged to
generate work. In fact, the output bias $b'(b, p_\text{max}, q_\text{max})$ for
an optimal ratchet is generally not equal to $1/2$. 

\subsection{Optimal memoryless ratchets versus memoryful ratchets}

At this point we may ask: Is it possible to surpass the optimal memoryless
ratchet in terms of work production with a memoryful ratchet? The answer seems
to be negative for memoryless inputs. More to the point,
Appendix~\ref{app:OptMemorylessInputs} proves the following statement: 
\begin{quote} 
For memoryless, binary inputs work production by the optimal memoryless ratchet cannot be surpassed by any memoryful ratchet.
\end{quote}
Thus, by optimizing over memoryless ratchets, we can actually determine the
optimum work production over all finite memoryful ratchets.
Appendix~\ref{app:OptMemorylessInputs} proves that for sequences of binary
symbols, memoryless ratchets are optimal for producing work. This has a number
of implications. First of all, it means that the dashed (blue) curve in Fig.
\ref{fig:MemorylessWork} is not only a bound on the work production of a
memoryless ratchet for any input with bias $b$, but it is also a bound on the
work production of \emph{any} finite memory ratchet with a memoryless input
with the same bias. In particular, the work production is at most $k_B T/e$, as
shown by the dashed (red) horizontal line. Importantly, this line is less than
the conventional Landauer bound of $k_B T \ln 2$. This seems counterintuitive,
since the Landauer bound on work production is often interpreted in the
literature to mean that as much as $k_B T \ln 2$ could be produced by
randomizing a totally predictable string. The lesson here, in contrast, is that
one must be careful invoking bounds, as they simply may not be achieved, and to
a substantial degree, for a large class of ratchets.

Appendix~\ref{app:OptMemorylessInputs}'s observation also suggests that
multiple ratchets in series---the output sequence of one is input to the
next---cannot be represented as a single finite-memory ratchet that interacts
with one bit at a time and only once. This is because we can surpass the work
production of an optimal memoryless ratchet with multiple ratchets interacting
with multiple symbols at a time, as we noted already. Ratchets composed in
series form a fundamentally different construction than a single memoryful
ratchet; a topic of some biological importance to which we will return
elsewhere.
 
\subsection{Infinite-memory ratchets}
\label{sec:InfiniteMemoryRatchets}

We emphasized that the very general IPSL bound on information processing based
on input-output entropy rate change holds for finite-state ratchets. What
happens if infinite memory is available to a ratchet? This section constructs
infinite-memory ratchets that can violate both Eqs.
(\ref{eq:EntropyRateSecondLaw}) and (\ref{eq:SingleSymSecondLaw}). The
intuition behind this is that, due to the infinite memory, the ratchet can
continue indefinitely to store information that need not be written to the
output. In effect, an apparent violation of the IPSL bound arises since the
hidden degrees of freedom of the ratchet's memory are not accounted for.

Nonetheless, infinite-memory ratchets offer intriguing possibilities for
thermodynamic and computational functionality. While finite-memory ratchets can
do meaningful computations and can even be appropriate models for, say,
biological organisms that have finite capacities and response times, they
cannot be computationally universal in the current architecture
\cite{Mins67,Lewi98a}. More precisely, one-way universal Turing machines
(UTMs), like our ratchet, read the input once and never again need an
``internal'' infinite work tape to read and write on. So, an infinite-state
ratchet of our type is needed to emulate the infinite bidirectional read-write
tape of the usual UTM~\cite{Broo89a}.

Appendix \ref{app:OptMemorylessInputs} showed that memoryless ratchets are able
to extract the most work from memoryless binary input processes, under the
assumption that the ratchet's memory is finite. Without finiteness the proof
breaks down, since an asymptotic state distribution may not exist over infinite
states \cite{Keme76a}. In addition, the proof of Eq.
(\ref{eq:EntropyRateSecondLaw}) fails for the same reason. Thus, we turn to
other tools for understanding the behavior in this case. The expression for
work production still holds, so despite not having general informational bounds
on work production, we can still calculate the exact work production for a
prototype infinite ratchet.

\begin{figure*}[tbp]
\centering
\includegraphics[width=1.8\columnwidth]{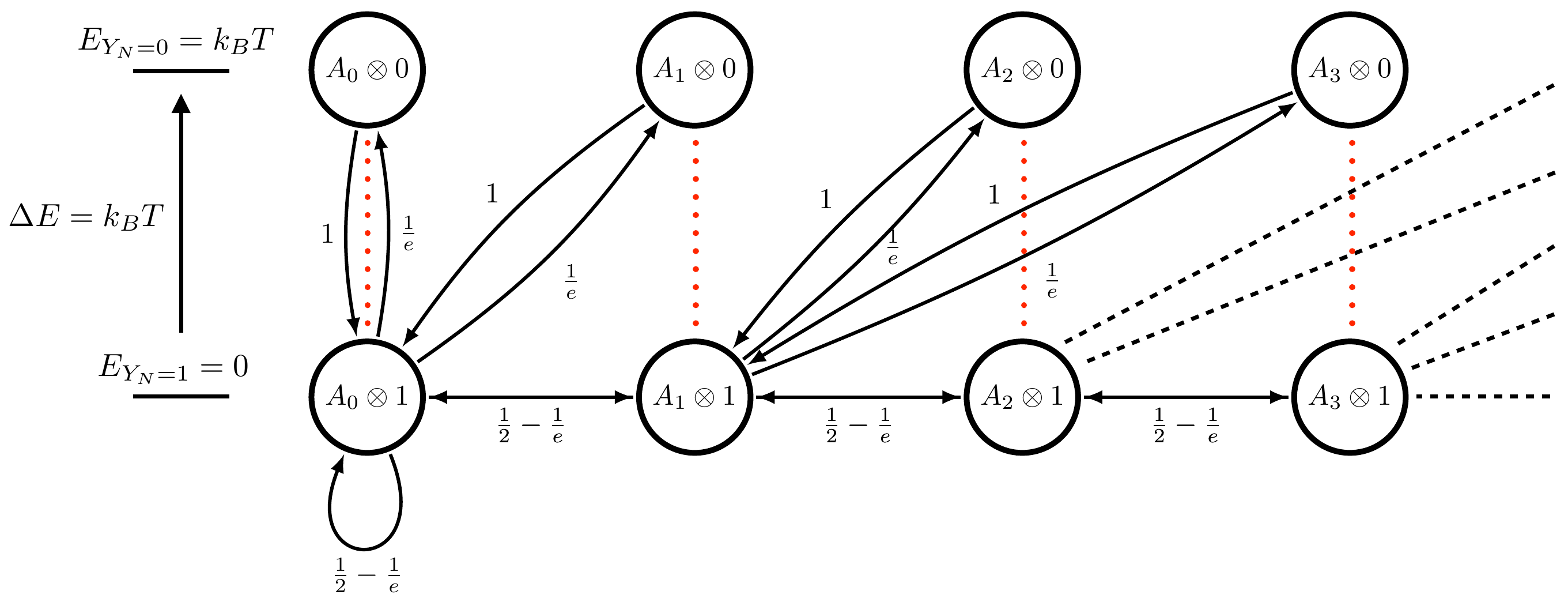}
\caption{Infinite-state ratchet that violates the IPSL and single-symbol bounds,
	Eqs.~(\ref{eq:EntropyRateSecondLaw}) and (\ref{eq:SingleSymSecondLaw}),
	respectively. The ratchet state-space is $\mathcal{X}=\{A_0, A_1, A_2,
	\ldots\}$: all states effectively have the same energy. The symbol values
	$\mathcal{Y}=\{0,1\}$ differ by energy $\Delta E = k_B T$, with $0$ having
	higher energy. The black arrows indicate the possible {\it interaction
	transitions} among the shown joint states of the ratchet and symbol
	during the interaction interval. For example, transitions $A_0 \otimes 1
	\leftrightarrow A_1 \otimes 1$ are allowed whereas transitions $A_0 \otimes
	0 \leftrightarrow A_1 \otimes 0$ are not. The dashed black lines show
	interaction transitions between the shown joint states and joint states
	that could not be shown. Briefly, for $i \geq 1$, there can be only the
	following interaction transitions: $A_i \otimes 1 \rightarrow \{A_{i \pm 1}
	\otimes 1, A_{2 i} \otimes 0, A_{2i+1} \otimes 0 \}$ and $ A_i \otimes 0
	\rightarrow A_{j(i)} \otimes 1$ with $j(i) = i/2$ for even $i$ and
	$(i-1)/2$ for odd $i$. For the $i = 0$ transitions, see the diagram.  Every
	interaction transition is followed by a switching transition and vice
	versa. The red dotted lines are the possible paths for driven
	\emph{switching transitions} between the joint states, which correspond to
	the production or dissipation of work. During the switching interval, the
	only allowed transitions are the vertical transitions between energy levels
	$A_i \otimes 0 \leftrightarrow A_i \otimes 1$.  The probability of these
	transitions depends on the input bias.
	}
\label{fig:InfiniteRatchet}
\end{figure*}

Here, we present an infinite-state ratchet with finite energy-differences
between all states. Our main result is that it produces more work than any
finite memory ratchet for a given input. More to the point, it violates both
the bounds in Eqs. (\ref{eq:EntropyRateSecondLaw}) and
(\ref{eq:SingleSymSecondLaw}). This demonstrates the need for the finite-memory
assumption in developing Landauer and IPSL bounds. Consider, for example, an
input process of all $1$s. According to
Sec.~\ref{sec:MemorylessRatchetEnergetics}, the maximum amount of work that can
be extracted from this input by a memoryless ratchet is given by:
\begin{align*}
\langle W \rangle_\text{max} = \frac{\kB T}{e}
  ~.
\end{align*}
The discussion in App.~\ref{app:OptMemorylessInputs} indicates that this should
be the maximum amount of work that can be extracted by any finite-memory
ratchet (for the same input). The infinite-state ratchet shown in Fig.
\ref{fig:InfiniteRatchet}, however, produces twice as much work:
\begin{align*}
\langle W \rangle^\infty = \frac{2 \kB T}{e}
  ~.
\end{align*}
The infinite-state ratchet also violates both of the IPSL and single-symbol
bounds, Eqs.~(\ref{eq:EntropyRateSecondLaw}) and (\ref{eq:SingleSymSecondLaw}),
since $k_B T \ln2$ is an upper bound for the work generation in all binary
input processes according to these bounds, whereas $2 / e > \ln 2$.

Let's describe the structure and dynamics of the infinite-state ratchet in
Fig.~\ref{fig:InfiniteRatchet} in detail. This ratchet has a countably infinite
number of states $A_i$, with $i \in \{0,1,2, \ldots\}$. In other words, the
ratchet state space is $\mathcal{X}=\{A_0,A_1,A_2, \ldots \}$. The joint
dynamic of the ratchet and the interacting symbol is shown in
Fig.~\ref{fig:InfiniteRatchet}, where the arrows indicate allowed transitions
and the number along the arrow, the associated transition probabilities. Apart
from the case $i = 0$, only the following transitions are allowed: $A_i \otimes
1 \rightarrow \{A_{i\pm1} \otimes 1, A_{i+1} \otimes 0 \}$ and $ A_i \otimes 0
\rightarrow A_{j} \otimes 1$ with $j = i/2$ for even $i$ and $(i-1)/2$ for odd
$i$. If the incoming symbol is $0$, the only transition allowed involves a
simultaneous change in the ratchet state and symbol, switching over to state
$A_{j(i)}$ if it started in state $A_i$ and the symbol switching to $1$. The
only exception is the case $i = 0$ in which the ratchet stays in the same
state, while the symbol switches to $1$. If the incoming symbol is 1, there are
generally three possible transitions: $A_i \otimes 1 \rightarrow A_{i \pm1}
\otimes 1$ and $A_i \otimes 1 \rightarrow A_{i + 1} \otimes 0$. The first two
transitions occur with equal probabilities $1/2 - 1/e$, while the third
transition occurs with probability $1/e$. For $i = 0$, there are four
transitions possible: $A_0 \otimes 1 \rightarrow \{A_0 \otimes 1 \,
\text{(self-loop)} ,  A_{1} \otimes 1, A_0 \otimes 0, A_1 \otimes 1 \}$. The
transition probabilities are shown in the figure.
	
We can assign relative energy levels for the joint states $A_i \otimes \{0,1\}$
based on the transition probabilities. Since the (horizontal) transitions $A_i
\otimes 1 \leftrightarrow A_{i+1} \otimes 1$ have equal forward and reverse
transition probabilities, all the joint states $A_i \otimes 1$ have the same
energy. Any state $A_i \otimes 0$ is higher than the state $A_{j(i) \otimes 1}$
by an energy:
\begin{align*}
\Delta E_{A_i \otimes 1 \rightarrow A_j \otimes 0} = k_B T \ln \frac{1}{1/e} = k_B T
  	~.
\end{align*}
As a result, all states $A_i \otimes 0$ have the same energy, higher than that
of the states $A_i \otimes 1$ by $k_B T$. This energy difference is responsible
for producing the work. When the ratchet is driven by the all $1$s process, if
it is in an $A_i \otimes 0$ state after the previous interaction transition,
then the switching transition changes the state to $A_i \otimes 1$ gaining
$\Delta E_{A_i \otimes 0 \rightarrow A_i \otimes1} = k_B T$ amount of work.
The probability of being in a $Y_N = 0$ state after an interaction interval is
$2/e$, so the work production is $\langle W \rangle = 2k_B T/e$, as stated
above.
	
The reason this infinite-state ratchet violates the information-theoretic
bounds is that those bounds ignore the asymptotic entropy production in the
ratchet's internal state space. There is no steady state over the infinite set
of states and this leads to the continual production of entropy within the
ratchet's state space $\mathcal{X}$. For the specific case of all $1$s input
process note that, before the interaction interval, the joint state-space
distribution of the ratchet and the incoming symbol must be positioned over
only $A_i \otimes 1$ states. This is due to the fact that the switching
transition always changes the symbol value to $1$. From a distribution $\{
\Pr(X_N=A_i,Y_N=1) \}_{i \in \{0,1,...\}}$ over the $A_i \otimes 1$ states at
time $N$, the interaction interval spreads the joint distribution to both $A_i
\otimes 0$ and $A_i \otimes 1$ states. However, they are reset to a new
distribution over the $A_i \otimes 1$ states $\{\Pr(X_{N+1}=A_i, Y_{N+1}=1)
\}_{i \in \{0,1,...\}}$ after the following switching transition. This leads to
a spreading of the probability distribution, and, therefore, to an increase in
entropy, in the ratchet space $\mathcal{X}$ after each time step.

\begin{figure}[tbp]
\centering
\includegraphics[width=\columnwidth]{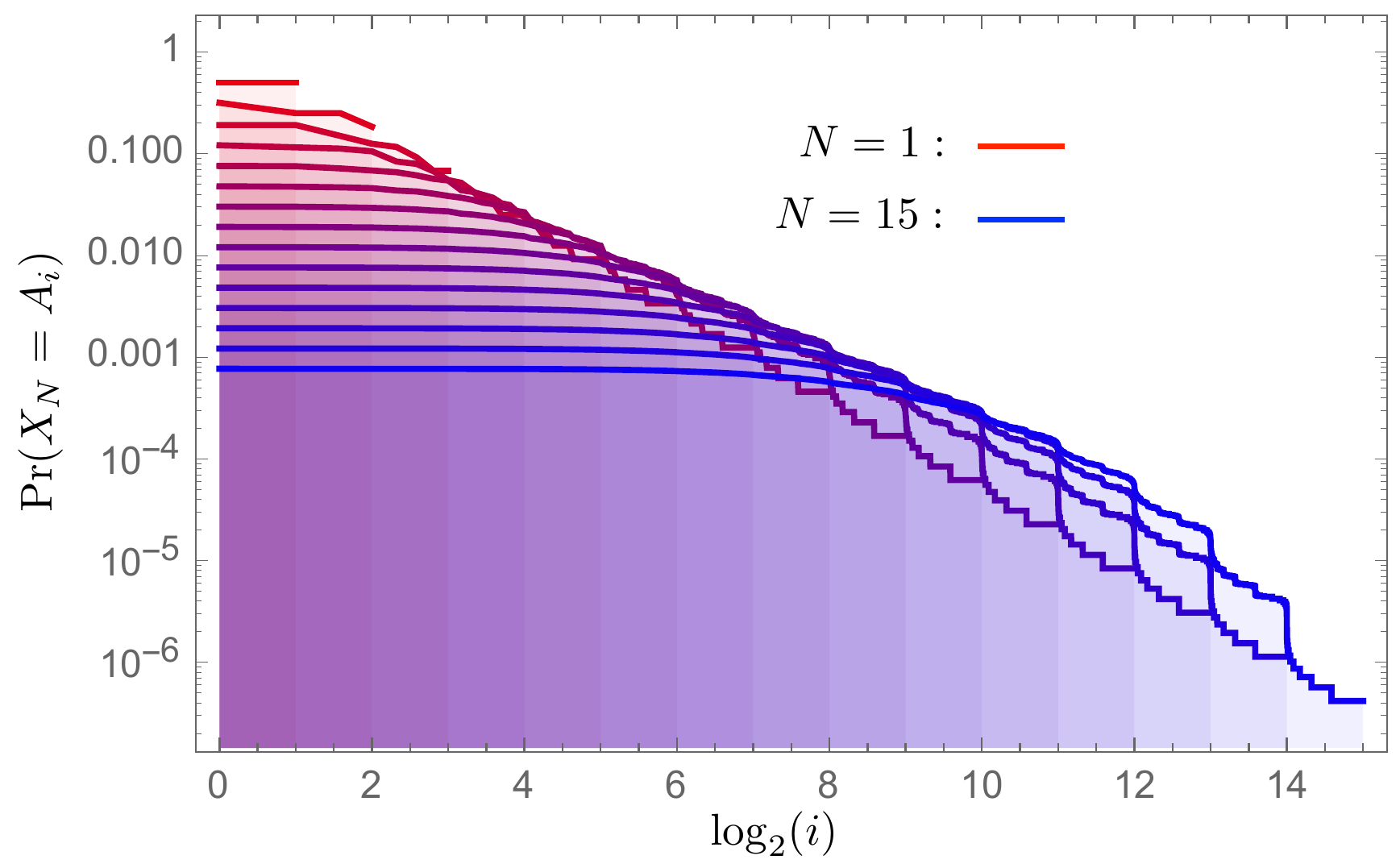}
\caption{Evolution of infinite-ratchet state distribution starting from an
	initial distribution peaked over the set $\mathcal{X}$, whose states
	are indexed $A_i$. The state distribution curves are plotted over $15$
	time steps, starting in red at time step $N=1$ and slowly turning to blue
	at time step $N=15$. With each sequential step, the support of the ratchet
	state distribution doubles in size, leading to increasing uncertainty in
	the ratchet state space and so in state entropy.
	}
\label{fig:Spreading}
\end{figure}

Figure \ref{fig:Spreading} demonstrates the spreading by setting the initial
joint ratchet-symbol state $X_0\otimes Y_0$ to $A_0\otimes 0$ and letting the
distribution evolve for $N=15$ time steps over the ratchet states. The ratchet
states are indexed by $i$ and the time steps are indexed by $N$, going from $1$
to $15$. The curves show the probabilities $\Pr(X_N=A_i)$ of the ratchet at
time step $N$ being in the $i$th ratchet state. By filling the area under each
distribution curve and plotting the ratchet-state index in logarithm base $2$,
we see that the distribution's support doubles in size after every time step.
This indicates an increase in the ratchet's internal entropy at each time step.
This increase in internal entropy is responsible for the violation of the IPSL
bounds in Eqs.~(\ref{eq:EntropyRateSecondLaw}) and
(\ref{eq:SingleSymSecondLaw}).
	
\begin{figure}[tbp]
\centering
\includegraphics[width=\columnwidth]{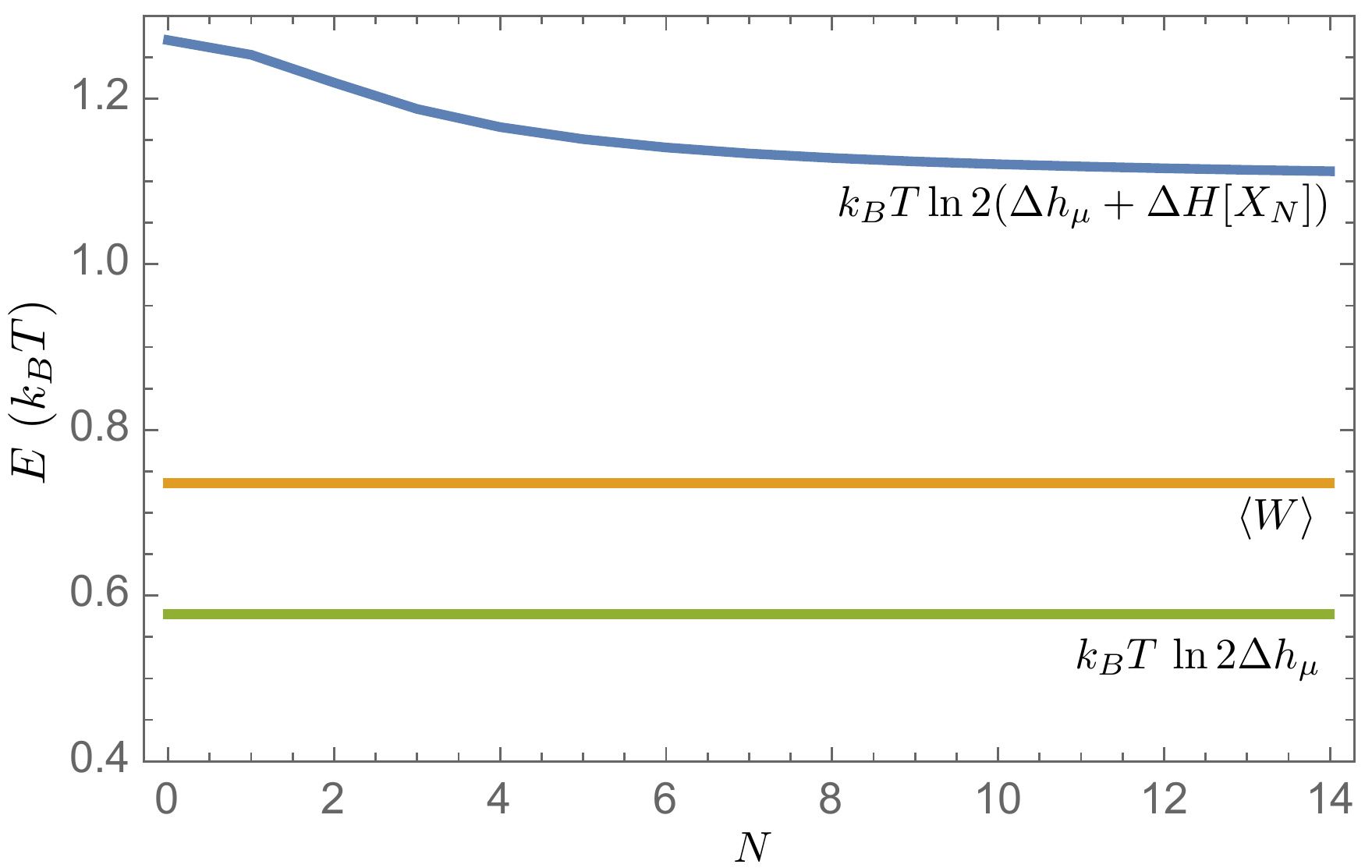}
\caption{The dashed (orange) line indicates average work production
	$\langle W \rangle$ per time step. It lies above the dotted (green) curve
	that indicates the IPSL entropy-rate bound on $\langle W \rangle$
	(Eq.~(\ref{eq:EntropyRateSecondLaw})), indicating a violation of the
	latter. The interpretation of the violation comes from the solid (blue)
	curve that indicates the joint entropy production of the input process and
	the ratchet together. We see a violation of the entropy-rate bound since
	there is continuous entropy production in the ratchet's (infinite) state
	space.
	}
\label{fig:SpreadingEntropy}
\end{figure}

We have yet to discover a functional form for a steady state that is
invariant---that maps to itself under one time-step. We made numerical
estimates of the ratchet's entropy production, though. From the distributions
shown in Fig.~\ref{fig:Spreading}, we calculated the ratchet's state entropies
at each time step $N$. The entropy production $\Delta \H[X_N] = \H[X_{N+1}] -
\H[X_{N}]$ at the $N$th step is shown in Fig.~\ref{fig:SpreadingEntropy}. We
see that the sum $\Delta \H[X_N] + \Delta h_\mu$ of the changes in ratchet
entropy and symbol entropy upper bounds the work production. Note that only the
$\Delta h_\mu$ curve lies below the work production. Thus, while this infinite
ratchet violates the IPSL bounds of Eqs. (\ref{eq:EntropyRateSecondLaw}) and
(\ref{eq:SingleSymSecondLaw}), it still satisfies a more general version of the
Second Law of Thermodynamics for information ratchets---Eq. (A7) of
Ref.~\cite{Boyd15a}:
\begin{align}
\langle W_N \rangle \leq k_B T \ln{2} \left( \H_{N+1} - \H_{N} \right)
  ~,
\label{eq:GenSecondLaw}
\end{align}
where $W_N$ is the work gain at the $N$th time step and $\H_N = \H[X_N,
Y_{N:\infty}, Y'_{0:N}]$ is the joint Shannon entropy of the ratchet and the
input and output symbol sequences $Y_{N:\infty}$ and $Y'_{0:N}$, respectively,
at time $t = N$. As we can see, this bound is based on not only the input and
output process statistics, but also the ratchet memory.

\section*{Conclusion}
\label{sec:Conclusion}

How an agent interacts with and leverages it's environment is a topic of broad
interest, from engineering and cybernetics to biology and now physics
\cite{Ashb57a,Ehre80}. General principles for how the structure of an agent
must match that of its environment will become essential tools for
understanding how to take thermodynamic advantage of correlations in structured
environments, whether the correlations are temporal or spatial. Ashby's Law of
Requisite Variety---a controller must have at least the same variety as its
input so that the whole system can adapt to and compensate that variety
\cite{Ashb57a}---was an early attempt at such a general principle of regulation
and control. In essence, a controller's variety should match that of its
environment. Above, paralleling this, we showed that a thermal agent
(information engine) interacting with a structured input (information
reservoir) obeys a similar variety-matching principle.

For an efficient finite-state information ratchet, the ratchet memory should
reflect the memory of the input process. More precisely, memoryless ratchets
are optimal for leveraging memoryless inputs, while memoryful ratchets are
optimal for leveraging memoryful inputs. This can be appreciated in a two
different ways.

On the one hand, the first comes from information processing properties of the
ratchet and input and the associated IPSL bounds on work. The operation of
memoryless ratchets can only destroy temporal correlations. These ratchets'
work production is still bounded by single-symbol entropy changes, as in
Eq.~(\ref{eq:SingleSymSecondLaw}). And, since memoryless input processes only
produce single-symbol correlations (statistical biases), the memoryless ratchet
bound of Eq.~(\ref{eq:SingleSymSecondLaw}) allows for maximal work production.
Thus, according their bounds, memoryless ratchets and inputs produce the most
work when paired.

On the other hand, in the second view memoryful input processes exhibit
multiple-symbol temporal correlations. And, the entropy rate bound of Eq.
(\ref{eq:EntropyRateSecondLaw}) suggests that the memoryful input processes can
be used to produce work in a memoryful ratchet, but not a memoryless one. More
precisely, we can conceive of memoryful input processes whose single-symbol
statistics are unbiased (equal proportions of $0$s and $1$s, in case of binary
alphabet) but the entropy rate is smaller than the single-symbol entropy:
$h_\mu < \H_1 ~(= \ln{2} \, \text{for a binary alphabet})$. In this case, since
the single-symbol entropy is already at its maximum possible value, memoryless
ratchets are unable to extract any work. Since the memoryful ratchets satisfy
the IPSL bound of Eq.~(\ref{eq:EntropyRateSecondLaw}), however, they can
extract work from such memoryful processes. One such example is studied in
detail by Ref.~\cite{Boyd16c}. For a quantum-mechanical ratchet, compare
Ref.~\cite{Chap15a}. Thus, memoryful ratchets are best paired with memoryful
inputs. This and its complement result---memoryless inputs are optimally used
by memoryless ratchets---is biologically suggestive. If one observes memory
(temporal correlations) in the transduction implemented by a biomolecular
assembly, for example, then it has adapted to some structured environment.
	
We summarized the role of memory in thermodynamic processes in
Fig.~\ref{tab:BoundTable} which considers each of the four possible
combinations of memoryful or memoryless ratchets with memoryful or memoryless
input.

While the Second Law of Thermodynamics determines the IPSL and related bounds
discussed here, it does not follow that the bounds are achievable for the class
of information ratchets considered. Based on an exact method for calculating
the average work production~\cite{Boyd16c}, we saw that there are indeed
situations where the bounds are not achievable (Fig.~\ref{fig:MemorylessWork}).
In Sec.~\ref{sec:MemorylessRatchetEnergetics}, we saw that memoryless ratchets
cannot generally saturate their bound (Eq.~(\ref{eq:SingleSymSecondLaw})).
Furthermore, based on the results of App. \ref{app:OptMemorylessInputs} we
could prove that finite memoryful ratchets fare no better than memoryless
ratchets at leveraging memoryless inputs. Thus, not even memoryful ratchets can
extract the maximum amount of work possible from a memoryless input. There are
some hints, though, as to what the architecture of information engines should
be to extract the maximum possible work allowed by the Second Law. We alluded
to one such situation in Sec.~\ref{sec:MemorylessRatchetEnergetics} involving a
``swarm'' of memoryless, optimized ratchets.

The unattainability of the IPSL bound observed above pertains to the
architecture of information engines where there is only a single ratchet that
interacts with one environmental signal value at a time. This leads one to
speculate that multiple ratchets interacting with different signals---say,
chained together so that the output of one is the input of another---will lead
to a closer approach to the bound. Simply having multiple copies of the optimal
memoryless ratchets one after another, however, will not necessarily address
unattainability. Interestingly, depending on input bias $b$, there may be
oscillations in the amount of work that is gained per cycle. And, even with
infinitely many ratchets chained together sequentially, we may still be far
from the IPSL bound. Based on our intuition about thermodynamically reversible
processes we postulate that to approach the bound more closely we need
increasingly many memoryless ratchets, each optimized with respect to \emph{its
own input}. We leave the verification of this intuition for a future
investigation. This does suggest, though, architectural trade-offs that should
manifest themselves in evolved biological thermodynamic processes.
	
To complete our exploration of the role of memory in thermodynamic processes,
we considered infinite-state ratchets, which are necessary if we wish to
physically implement universal Turing machines with the unidirectional
information ratchets. Infinite ratchets, however, pose a fundamental challenge
since the IPSL entropy-rate bound on work production does not apply to them.
The proof of the bound (Eq.~(\ref{eq:EntropyRateSecondLaw})~\cite{Boyd16c}) is
based on the assumption that the ratchet reaches a steady state after
interacting with sufficiently many input symbols. This need not be the case for
infinite-state ratchets. In fact, the numerical investigations of
Sec.~\ref{sec:InfiniteMemoryRatchets} indicate that the probability
distribution in the state space of an infinite ratchet can continue to spread
indefinitely, without any sign of relaxing to a steady state; recall
Fig.~\ref{fig:Spreading}. By calculating both the average work production per
time step and the amount of change in the entropy rate,
Fig.~\ref{fig:SpreadingEntropy} showed that there is a violation of the IPSL
and related bounds. This necessitates a modification of the IPSL for
infinite-state ratchets. The appropriate bound, though, has already been
presented in a previous work~\cite{Boyd16c}, which we quoted in
Eq.~(\ref{eq:GenSecondLaw}). This relation shows that the work production is
still bounded by the system's entropy production; only, we must include the
contribution from the ratchet's internal state space on top of the entropy-rate
difference of the input and the output HMMs.
	
We close by highlighting the close correspondence between information ratchets
and biological enzymes. Most directly, it is possible to model the biomimetic
enzymes following the design of information ratchets ~\cite{Cao15}. The
correspondence goes further, though. In Sec.~\ref{sec:Synopsis}, we discussed
how a swarm of ratchets acting cooperatively may be more efficient than
individual information ratchets, even if they are quite sophisticated. A
similar phenomenon holds for enzymes where the enzymes along a metabolic
pathway assemble to form a multi-enzyme complex---a ``swarm''---to affect
faster, efficient reaction turnover, known as \emph{substrate
channeling}~\cite{Phil08}.

\section*{Acknowledgments}

As an External Faculty member, JPC thanks the Santa Fe Institute for its
hospitality during visits. This work was supported in part by the U. S. Army
Research Laboratory and the U. S. Army Research Office under contracts
W911NF-13-1-0390 and W911NF-12-1-0234.

\appendix
 
\section{Optimally Leveraging Memoryless Inputs}
\label{app:OptMemorylessInputs}

It is intuitively appealing to think that memoryless inputs are best utilized
by memoryless ratchets. In other words, the optimal ratchet for a memoryless
input is a memoryless ratchet. We prove the validity of this intuition in the
following. We start with the expression of work production per time step:
\begin{align*}
\beta \langle W \rangle &= \sum_{x,x',y,y'} \pi_{x \otimes y} M_{x \otimes y \rightarrow x' \otimes y'} \ln \frac{M_{x' \otimes y' \rightarrow x \otimes y}}{M_{x \otimes y \rightarrow x' \otimes y'}}
\\&= \sum_{x,x',y,y'} \pi_{x \otimes y} M_{x \otimes y \rightarrow x' \otimes y'}\ln \frac{\pi_{x' \otimes y'}M_{x' \otimes y' \rightarrow x \otimes y}}{\pi_{x \otimes y}M_{x \otimes y \rightarrow x' \otimes y'}} \nonumber
\\& -\sum_{x,x',y,y'} \pi_{x \otimes y} M_{x \otimes y \rightarrow x' \otimes y'}\ln \frac{\pi_{x' \otimes y'}}{\pi_{x \otimes y}}, \nonumber
\end{align*}
with $\beta=1/k_B T$. The benefit of the decomposition in the second line will
be clear in the following. Let us introduce several quantities that will also
be useful in the following:
\begin{align*}
p(x,y,x',y') & =\pi_{x \otimes y} M_{x \otimes y \rightarrow x' \otimes y'}
	~, \\
	p_R(x,y,x',y')&  = \pi_{x' \otimes y'} M_{x' \otimes y' \rightarrow x
	\otimes y}  ~, \\
	\pi^X_x & =\sum_y \pi_{x \otimes y}  ~, \\
	\pi^Y_y & =\sum_x \pi_{x \otimes y}  ~, \\
	p^X(x,x') & =\sum_{y,y'}p(x,y,x',y')  ~,~\text{and}\\
	p^Y(y,y') & =\sum_{x,x'}p(x,y,x',y').
\end{align*}
For a memoryless input process, sequential inputs are statistically
independent. This implies $Y_N$ and $X_N$ are independent, so the stationary
distribution $\pi_{x \otimes y}$ can be written as a product of marginals:
\begin{align}
\pi_{x \otimes y} = \pi^X_x \pi^Y_y.
\end{align}
In terms of the above quantities, we can rewrite work for a memoryless input
process as:
\begin{align*}
\beta \langle W \rangle & = D_{KL}(p||p_R) \\
  & \quad\quad - \sum_{y,y'}p^Y(y,y')\ln \frac{\pi^Y_{y'}}{\pi^Y_y}
  -\sum_{x,x'}p^X(x,x')\ln \frac{\pi^X_{x'}}{\pi^X_x} 
  ~,
\end{align*}
where $D_{KL}(p||p_R)$ is the relative entropy of the distribution $p$ with
respect to $p_R$ \cite{Cove06a}. Note that the last term in the expression
vanishes, since the ratchet state distribution is the same before and after an
interaction interval:
\begin{align}
\sum_x p^X(x,x') =\sum_x p^X(x',x) =\pi^X_{x'},
\end{align}
and so:
\begin{align*}
\sum_{x,x'} & p^X(x,x') \ln \frac{\pi^X_{x'}}{\pi^X_x} \\
   & = \sum_{x,x'}p^X(x,x')\ln \pi^X_{x'}-\sum_{x,x'}p^X(x,x')\ln \pi^X_{x} \\
   & = \sum_{x'}\pi^X_{x'}\ln \pi^X_{x'}-\sum_{x}\pi^X_{x}\ln \pi^X_{x}
\\ & = 0
  ~.
\end{align*}
Thus, we find find the average work production to be:
\begin{align}
\beta \langle W \rangle =-D_{KL}(p||p_R)-\sum_{y,y'}p^Y(y,y')
  \ln \frac{\pi^Y_{y'}}{\pi^Y_y}
  ~.
\end{align}

Let us now use the fact that the coarse graining of any two distributions, say
$p$ and $q$, yields a smaller relative entropy between the two~\cite{Cove06a,
Gome08a}. In the work formula, $p^Y$ is a coarse graining of $p$ and $p^Y_R$ is
a coarse graining of $p^R$, implying:
\begin{align}
D_{KL}(p^Y||p^Y_R) \leq D_{KL}(p||p_R)
  ~.
\end{align}
Combining the above relations, we find the inequality:
\begin{align*}
\beta \langle W \rangle
  \leq -D_{KL}(p^Y||p^Y_R)-\sum_{y,y'}p^Y(y,y')\ln \frac{\pi^Y_{y'}}{\pi^Y_y}
  ~.
\end{align*}

Now, the marginal transition probability $p^Y(y,y')$ can be broken into the
product of the stationary distribution over the input variable $\pi^Y_{y}$ and
a Markov transition matrix $M^Y_{y \rightarrow y'}$ over the input alphabet:
\begin{align*}
p^Y(y,y') = \pi^Y_yM^Y_{y \rightarrow y'}
  ~,
\end{align*}
which for any ratchet $M$ is:
\begin{align*}
M^Y_{y \rightarrow y'} & = \frac{1}{\pi_y^Y} p^Y(y, y') \\
& = \frac{1}{\pi_y^Y} \sum_{x,x'} \pi_{x \otimes y} M_{x \otimes y \rightarrow x' \otimes y'} \\
& = \frac{1}{\pi_y^Y} \sum_{x,x'} \pi_x^X \pi_y^Y M_{x \otimes y \rightarrow x' \otimes y'} \\
& = \sum_{x,x'} \pi_x^X M_{x \otimes y \rightarrow x' \otimes y'}
  ~.
\end{align*}
We can treat the Markov matrix $M^Y$ as corresponding to a ratchet in the same
way as $M$. Note that $M^Y$ is effectively a memoryless ratchet since we do not
need to refer to the internal states of the corresponding ratchet. See
Fig.~\ref{fig:MemoryDifferences}. The resulting work production for this
ratchet $\langle W^Y \rangle$ can be expressed as:
\begin{align*}
\beta \langle W^Y \rangle & =\sum_{y,y'} \pi^{Y}_y M^Y_{y \rightarrow y'}
  \ln\frac{M^Y_{y' \rightarrow y}}{M^Y_{y \rightarrow y'}} \\
  & = -D_{KL}(p^Y||p^Y_R)-\sum_{y,y'}p^Y(y,y')\ln \frac{\pi^Y_{y'}}{\pi^Y_y} \\
  & \geq \beta \langle W \rangle
  ~.
\end{align*}
Thus, for any memoryful ratchet driven by a memoryless input we can design a
memoryless ratchet that extracts at least as much work as the memoryful
ratchet.

There is, however, a small caveat. Strictly speaking, we must assume the case
of binary input. This is due to the requirement that the matrix $M$ be detailed
balanced (see Sec.~\ref{sec:Synopsis}) so that the expression of work used here
is reliable. More technically, the problem is that we do not yet have a proof
that if $M$ is detailed balanced then so is $M^Y$, a critical requirement
above. In fact, there are examples where $M^Y$ does not exhibit detailed
balance. We do, however, know that $M^Y$ is guaranteed to be detailed balanced
if $\mathcal{Y}$ is binary, since that means $M^Y$ only has two states and all
flows must be balanced. Thus, for memoryless binary input processes, we
established that there is little point in using finite memoryful ratchets to
extract work: memoryless ratchets extract work optimally from memoryless binary
inputs.

\bibliography{chaos}

\end{document}